\documentclass[prd, preprint, tightenlines, superscriptaddress, showpacs, byrevtex]{revtex4-1}

\usepackage{blindtext}
\usepackage{graphicx}
\usepackage{subfigure}
\usepackage{amsmath}
\usepackage{multirow}
\usepackage{mathtools}
\usepackage{overpic}
\usepackage{epsfig}
\usepackage{longtable}
\usepackage{dcolumn}
\usepackage{bm}
\usepackage{color}
\usepackage{siunitx}
\usepackage{lineno}
\usepackage[figuresright]{rotating}
\usepackage{diagbox}
\usepackage[normalem]{ulem}

\newcommand\redout{\bgroup\markoverwith
{\textcolor{red}{\rule[.5ex]{2pt}{0.4pt}}}\ULon}
\usepackage{xcolor}
\usepackage{threeparttable}

\newcommand{\gev}{\rm GeV}

\newcommand{\gevcs}{{\rm GeV}/c^2}
\newcommand{\mev}{\rm MeV}
\newcommand{\mevcs}{{\rm MeV}/c^2}

\newcommand{\ev}{\rm eV}

\newcommand{\pb}{\si{\pico\barn}}
\newcommand{\fb}{\si{\femto\barn}}
\newcommand{\infb}{\fb^{-1}}
\newcommand{\inpb}{\pb^{-1}}

\newcommand{\BR}{{\cal B}}

\newcommand{\piz}{\pi^0}

\newcommand{\ktb}{\bar{K}^{*}(892)}

\newcommand{\etap}{\eta^{\prime}}

\newcommand{\psp}{\psi(2S)}

\newcommand{\jpsi}{J/\psi}

\newcommand{\phit}{\phi(1680)}
\newcommand{\phiy}{\phi(2170)}
\newcommand{\EE}{e^+e^-}

\newcommand{\kkt}{K\ktb}
\newcommand{\GG}{\gamma\gamma}

\newcommand{\pp}{\pi^+\pi^-}
\newcommand{\ppp}{\pi^+\pi^-\pi^0}
\newcommand{\kk}{K^+K^-}

\newcommand{\ssb}{s\bar{s}}

\newcommand{\reduline}{\bgroup\markoverwith
{\textcolor{red}{\rule[0.5ex]{2pt}{0.4pt}}}\ULon}

\newcommand{\beq}{\begin{equation}}
\newcommand{\eeq}{\end{equation}}
\newcommand{\beqar}{\begin{eqnarray}}
\newcommand{\eeqar}{\end{eqnarray}}
\newcommand{\bitm}{\begin{itemize}}
\newcommand{\eitm}{\end{itemize}}

\def\Journal#1#2#3#4{{#1} {\bf #2}, #3 (#4)}
\def\IJMP{Int. J. Mod. Phys. A}

\def\PLB{Phys. Lett. B}
\def\PRL{Phys. Rev. Lett.}
\def\PRD{Phys. Rev. D}

\def\CPC{Chin. Phys. C}
\def\CoPC{Comput. Phys. Commun.}

\def\PTEP{ Prog. Theor. Exp. Phys. }
\def\PS{Physica Scripta}

\begin{document}

\title{
\quad\\[1.0cm]
Determination of the resonant parameters of excited vector strangenia with the $\EE\to\eta\phi$ data}

\author{Wenjing Zhu}\affiliation{Key Laboratory of Nuclear Physics and Ion-beam Application, Ministry of Education, 
China}
\affiliation{Institute of Modern Physics, Fudan University, Shanghai 200443}
\author{Xiaolong Wang}\affiliation{Key Laboratory of Nuclear Physics and Ion-beam Application, Ministry of Education, 
China}
\affiliation{Institute of Modern Physics, Fudan University, Shanghai 200443} 

\date{\today}

\begin{abstract}

We determine the resonant parameters of the vector states $\phit$ and $\phiy$ by doing a combined fit to the $\EE\to 
\eta\phi$ cross sections from threshold to $2.85~\gev$ measured by BaBar, Belle, BESIII and CMD-3 experiments. The 
mass $(1678 ^{+5}_{-3} \pm 7)~\mevcs$ and the width $(156 \pm 5 \pm 9)~\mev$ are obtained for the $\phit$, and the 
mass $(2169 \pm 5 \pm 6)~\mevcs$ and the width $(96^{+17}_{-14} \pm 9)~\mev$ for the $\phiy$. The statistical 
significance of $\phiy$ is $7.2\sigma$. Depending on the interference between the $\phit$, $\phiy$ and a non-resonant 
$\eta\phi$ amplitude in the nominal fit, we obtain four solutions and $\Gamma^{\EE}_{\phit}\cdot \BR[\phit\to\eta\phi]
= (79 \pm 4 \pm 16)$, $(127\pm 5 \pm 12)$, $(65^{+5}_{-4} \pm 13)$ or $(215^{+8}_{-5} \pm 11)~\ev$, and 
$\Gamma^{\EE}_{\phiy}\cdot \BR[\phiy\to\eta\phi] = (0.56^{+0.03}_{-0.02}\pm 0.07)$, $(0.36^{+0.05}_{-0.03}\pm 0.07)$, 
$(38 \pm 1 \pm 5)$ or $(41 \pm 2 \pm 6)~\ev$, respectively. We also search for the production of $X(1750)\to\eta\phi$ 
and the significance is only $2.0\sigma$, then we determine the upper limit of $\Gamma^{\EE}_{X(1750)}\cdot 
\BR[X(1750) \to \eta\phi]$ at 90\% confidence level.

\end{abstract}


\maketitle

\section{Introduction}

Hadronic transitions with $\pp$ or $\eta$ emittance have contributed significantly to the discoveries of 
quarkonium(-like) states, such as the $Y(4260)$ in $\EE\to \pp\jpsi$ via initial-state radiation (ISR) by the BaBar
experiment~\cite{babay4260}. In searching for an $s\bar{s}$ version of the $Y(4260)$, BaBar discovered the $Y(2175)$
(now called `$\phiy$') in $\EE\to\pp\phi$ via ISR~\cite{y2175_babar}, and later Belle confirmed it~\cite{y2175_belle}.
In searching for $\phiy$ in the hadronic transition with $\eta$, BaBar studied the $\EE\to\eta\phi$ process via ISR 
using a $232~\infb$ data sample and found an excess with the mass of $(2125\pm 22\pm 10)~\mevcs$ (tens $\mevcs$ lower 
than the world average value of $\phiy$~\cite{PDG}) and the width of $(61\pm 50\pm 13)~\mev$~\cite{etaphi_babar, 
etaphi_babar_2}. Hereinafter, the first quoted uncertainties are statistical and the second ones are systematic. 
Belle measured this process with much larger statistics in a $980~\infb$ data sample, but did not find this excess, 
and the statistical significance $\phiy$ is only $1.7\sigma$~\cite{etaphi_belle}. 

There are interesting measurements from the CMD-3 experiment and the BESIII experiment in the past years. The CMD-3 
experiment measured the process $\EE\to \kk\eta$ from 1.59 to $2.007~\gev$ and found it is dominated by the 
$\eta\phi$ contribution~\cite{etaphi_cmd3}. CMD-3 then calculated the contribution to the anomalous magnetic moment 
of muon: $a_\mu^{\eta\phi}(E<1.8~\gev) = (0.321\pm 0.015\pm 0.016)\times 10^{-10}$, $a_\mu^{\eta\phi} (E<2.0~\gev) = 
(0.440\pm 0.015\pm 0.022)\times 10^{-10}$. With a $715~\inpb$ data sample taken at 22 CM energy points in the range 
between 2.00 and $3.08~\gev$, BESIII measured the Born cross sections of $\EE\to\eta\phi$~\cite{etaphi_bes3} and 
$\EE\to \phi\etap$ ~\cite{etapphi_bes3}. BESIII reported the observation of $\phiy$ in the $\eta\phi$ final state and 
determined its resonant parameters to be $m_{\phiy} = (2163.5 \pm 6.2 \pm 3.0)~\mevcs$ and $\Gamma_{\phiy} = 
(31.1^{+21.1}_{-11.6}\pm 1.1)~\mev$~\cite{etaphi_bes3}, in which the width is much narrower than the world average 
value of about $100~\mev$~\cite{PDG}. BESIII also observed a resonance near $2.17~\gevcs$ in the $\phi\etap$ final 
state with a statistical significance exceeding $10\sigma$~\cite{etapphi_bes3}. Assuming it is $\phiy$, one can infer 
the ratio $\BR[\phiy\to\phi\eta]/\BR[\phiy\to\phi\etap] = (0.23\pm 0.10 \pm 0.18)$, which is smaller than the 
prediction of $s\bar{s}g$ hybrid models by several orders of magnitude.

It is a puzzle that the $\phiy$ is not significant in the $\eta$ transition comparing with the $\pp$ transition, and 
the measurement of $\phiy$ in $\eta\phi$ final state is still poor. The lineshape of $\sigma(\EE\to\eta\phi)$ is much 
different from that of $\sigma(\EE\to\pp\phi)$~\cite{y2175_babar, y2175_belle}, which could be helpful for 
understanding the difference between $\EE\to\eta\jpsi$ and $\EE\to\pp\jpsi$. In $c\bar{c}$ sector, $\sigma(\EE \to 
\eta\jpsi)/\sigma(\EE\to\pp\jpsi) \approx 1$ at the peak of the $Y(4260)$, while $\sigma(\EE\to\eta\phi)/\sigma(\EE 
\to\pp\phi) \gg 1$ at the peak of $\phit$ and $\ll 1$ at the peak of $\phiy$ in the $s\bar{s}$ sector. In a recent 
lattice Quantum Chromodynamics (QCD) calculation~\cite{lattice-qcd}, the properties of the lowest two states comply 
with those of $\phi$ and $\phit$, but with no obvious correspondence to the $\phiy$. 

Besides the $\phit$, there is one more state called `$X(1750)$' as a candidate of $\ssb$ quarkonium. The observation 
of the $\phit$ in $\kk$ and $\kkt$ is sometimes cited as evidence that this state is an $\ssb$ quarkonium, as the 
radial excitation of $\phi$. But it was argued that one true evidence for $\phi$ as an $\ssb$ state should be the 
large branching fractions to hidden strangeness modes such as $\eta\phi$~\cite{strange}. The FOCUS experiment 
reported a high-statistics study of diffraction photo-production of $\kk$, and observed the $X(1750)$ with mass 
$(1753.5 \pm 1.5 \pm 2.3)~\mevcs$ and width $(122.3 \pm 6.2 \pm 0.8)~\mev$~\cite{FOCUS}. Meanwhile, FOCUS saw a 
slight enhancement below the $\phit$ region but no obvious $X(1750)$ signal in the $\kkt$ final state. 

If $\phit$ and $X(1750)$ are the same state, the mass measured in $\EE$ collision and photoproduction experiments 
typically has a difference of $50-100~\mevcs$, with $\kkt$ dominance in $\EE$ collision and $\kk$ dominance in 
photo-production. This may constitute evidence for two distinct states, although interference with 
$q\bar{q}~(q = u,~d)$ vectors may complicate a comparison of these two processes. This issue can be addressed by 
studying channels in which interference with $q\bar{q}~(q = u,~d)$ vectors is expected to be unimportant, notably 
$\eta\phi$. With a sample of 4.48 million $\psp$ events, BESIII performed the first partial wave analysis of $\psp 
\to \kk\eta$ and got the simultaneous observation of the $\phit$ and $X(1750)$ in the $\kk$ mass 
spectrum~\cite{kketa}, which indicates that the $X(1750)$ is distinct from the $\phit$. Meanwhile, BESIII determined 
the $X(1750)$ to be an $1^{--}$ resonance.

Since the cross sections of $\EE\to\eta\phi$ have been measured well by the BaBar, Belle, BESIII and CMD-3 
experiments, it is much helpful to consider all of them for better understanding on $\phit$, $X(1750)$ and $\phiy$. 
In this paper, we combine the measured $\sigma(\EE\to\eta\phi)$ from BaBar, Belle, BESIII and CMD-3 experiments to 
have a better precision of the lineshape, which is helpful for the study of the anomalous magnetic moment of muon. 
Then, we perform combined fits to these measured cross sections for the resonant parameters of the $\phit$ and 
$\phiy$, and estimate the production of $X(1750)$ in the $\eta\phi$ final state.

\section{Measurements of $\sigma(\EE\to\eta\phi)$ }

\begin{figure}[htbp]
\psfig{file=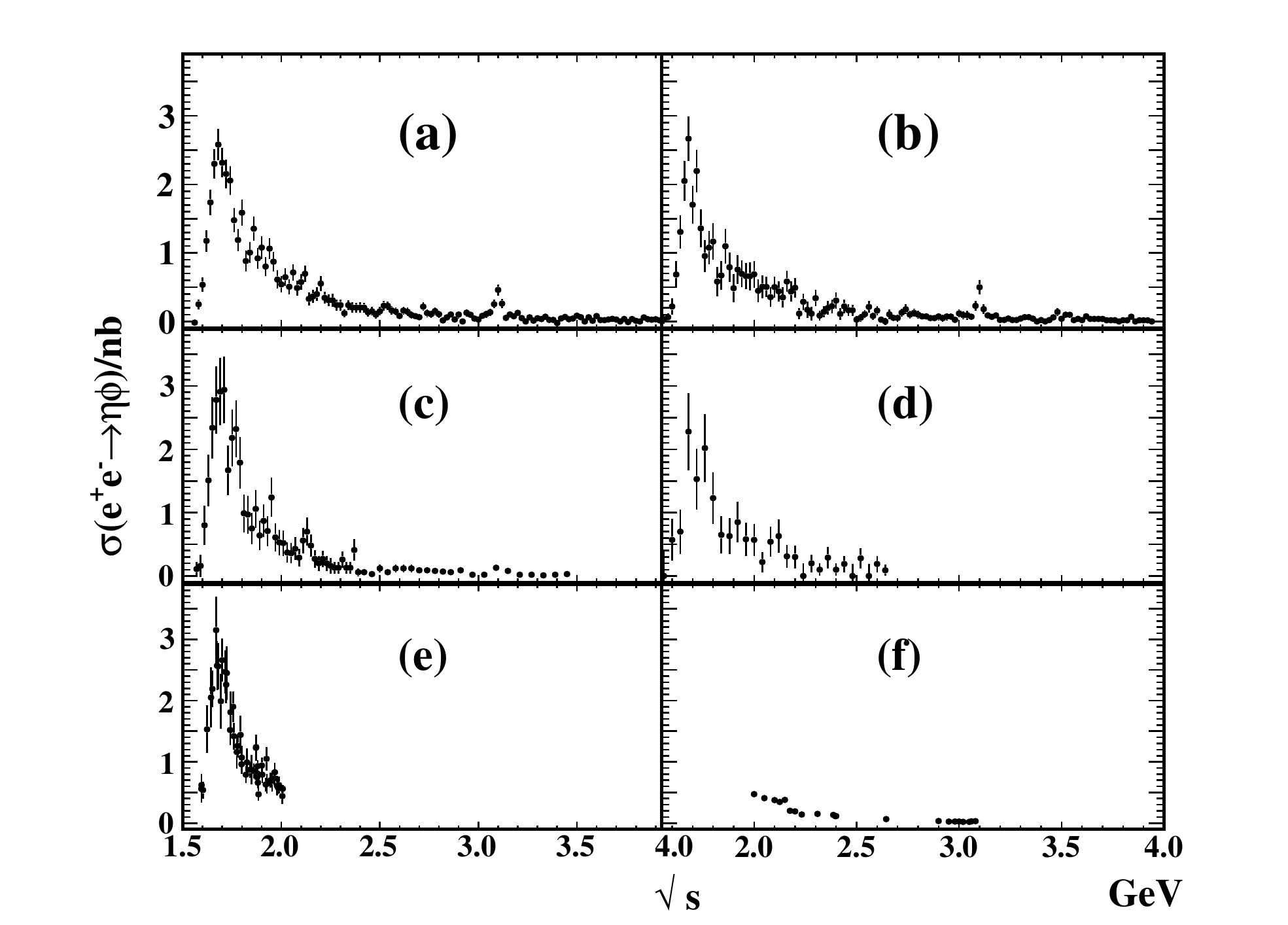,width=0.8\textwidth}
\caption{The $\sigma(\EE\to\eta\phi)$ measured in (a) the $\eta \to \GG$ mode at Belle, (b) the $\eta \to \ppp$ mode 
at Belle, (c) the $\eta \to \GG$ mode at BaBar, (d) the $\eta \to \ppp$ mode at BaBar, (e) the $\eta\to \GG$ mode at 
CMD-3 and (f) the $\eta \to \GG$ mode at BESIII.}
\label{xs_four_exps}
\end{figure}

The measurements of $\sigma(\EE\to\eta\phi)$ from the BaBar, Belle, BESIII and CMD-3 experiments are shown in 
Fig.~\ref{xs_four_exps}, in which from (a) to (f) are the Belle measurement with $\eta\to\GG$ 
mode~\cite{etaphi_babar}, Belle measurement with $\eta\to\ppp$~\cite{etaphi_babar_2}, BaBar measurement with $\eta\to 
\GG$ mode, BaBar measurement with $\eta\to\ppp$ mode~\cite{etaphi_belle}, CMD-3 measurement~\cite{etaphi_cmd3} and 
BESIII measurement~\cite{etaphi_bes3}, respectively. 

\bitm

\begin{figure}[tbp]
\includegraphics[width=0.4\textwidth]{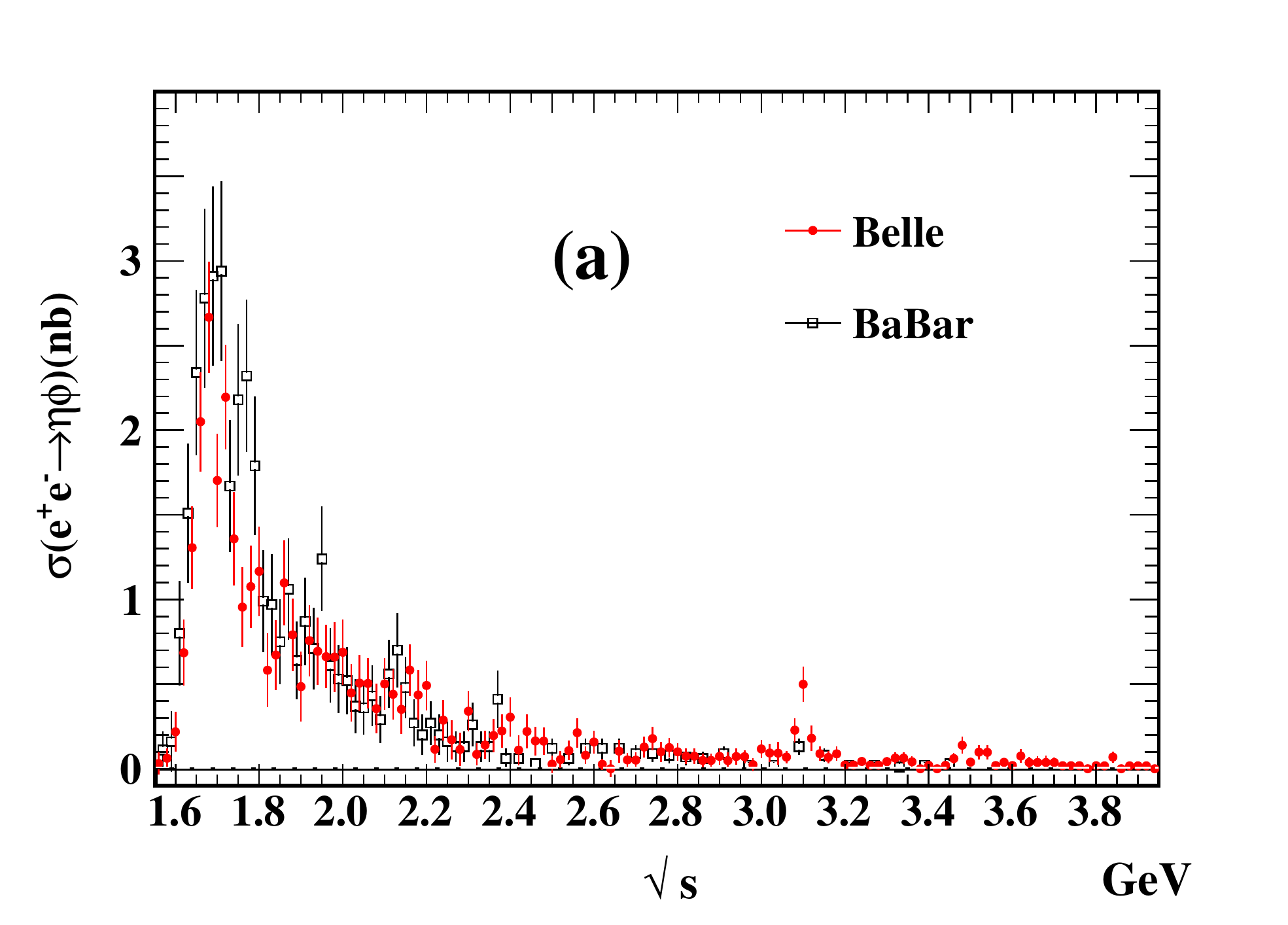}
\includegraphics[width=0.4\textwidth]{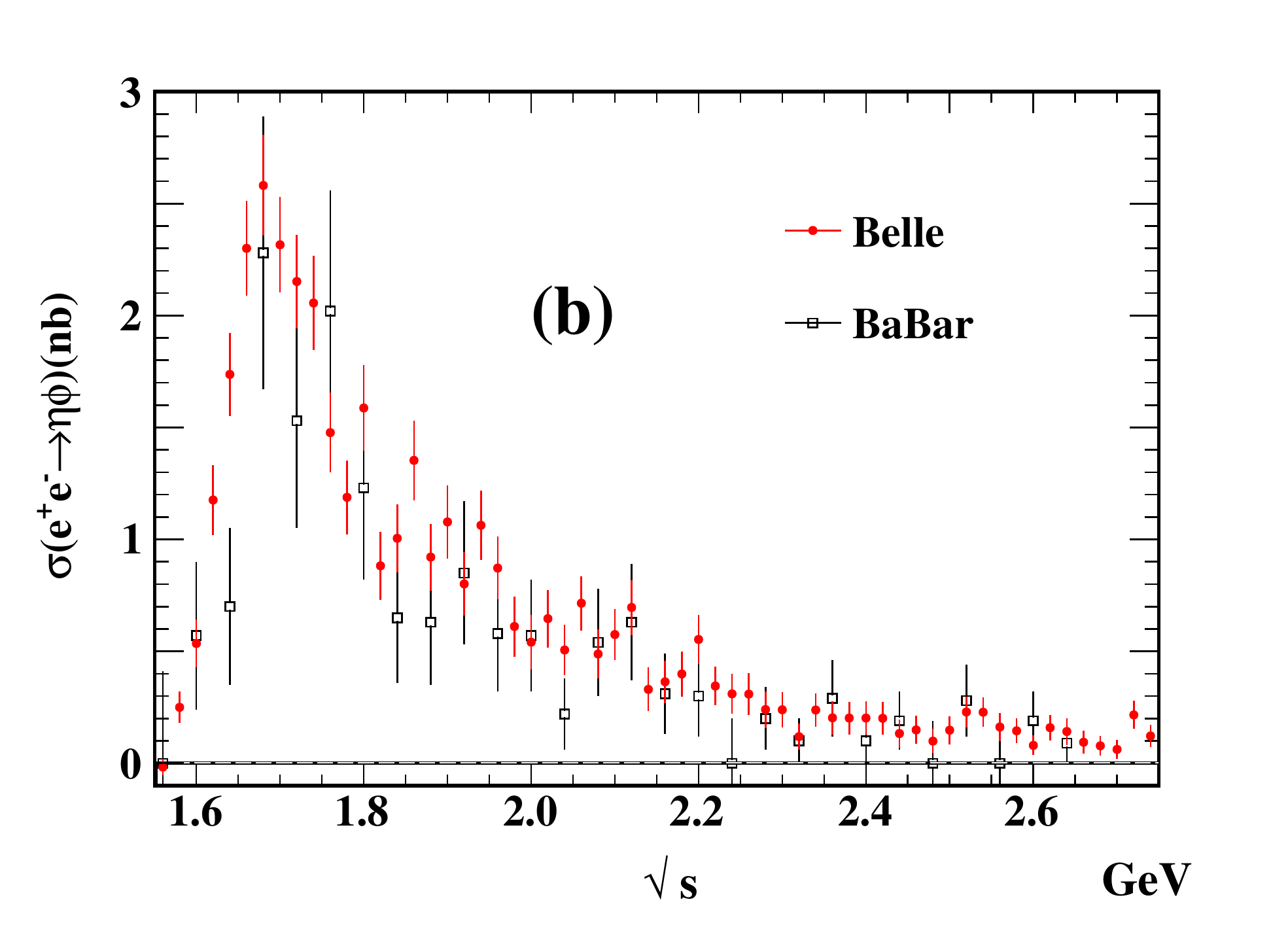}\\
\includegraphics[width=0.4\textwidth]{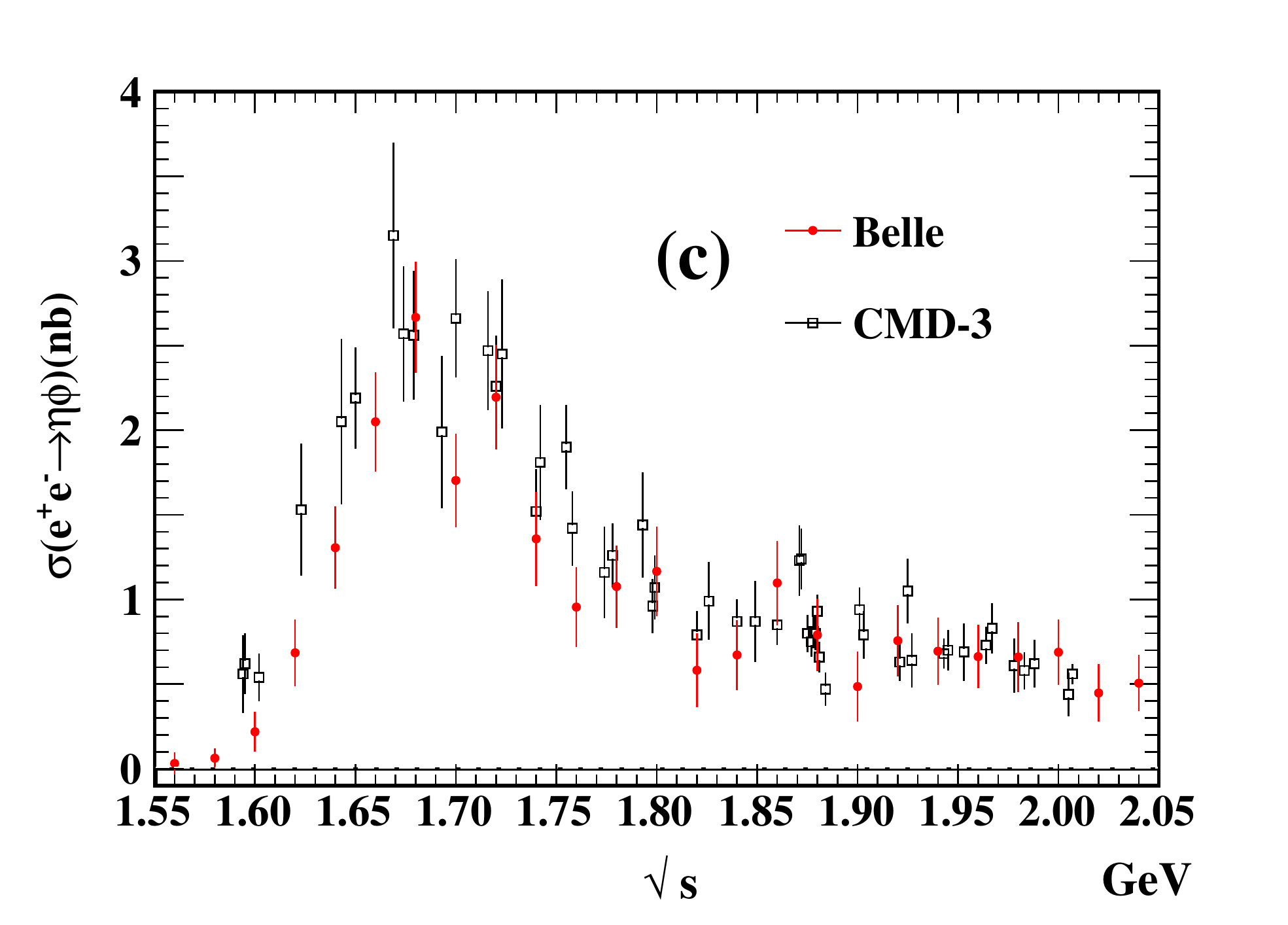}
\includegraphics[width=0.4\textwidth]{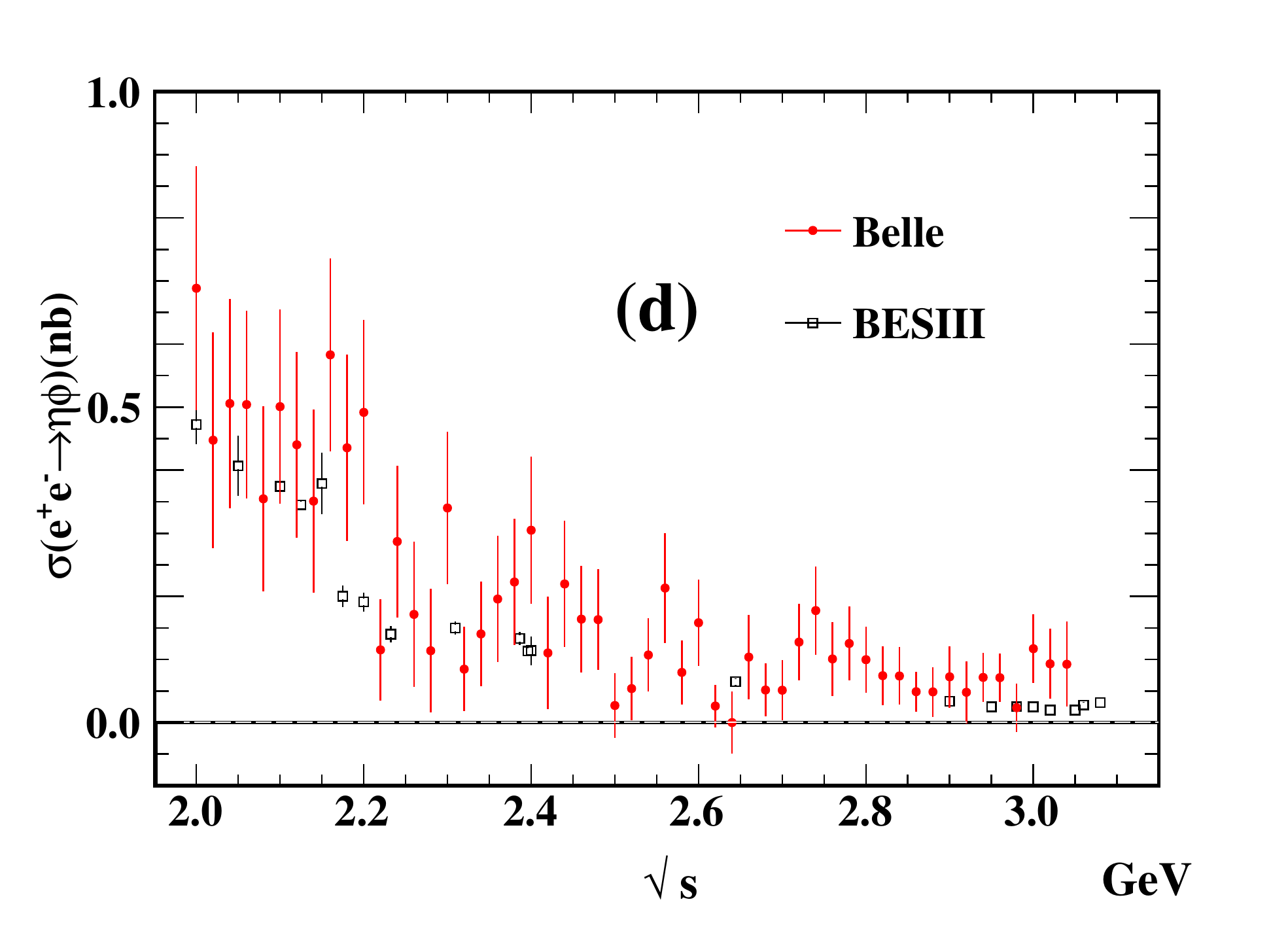}
\caption{The measurements on $\sigma(\EE\to\eta\phi)$ from the BaBar, BESIII and CMD-3 experiments comparing to the 
latest measurements from the Belle experiment. Plots (a) and (b) show the comparisons between BaBar and Belle in the 
$\eta\to \GG$ mode and the $\eta\to\pp\piz$ mode, respectively; plots (c) and (d) show the comparison between CMD-3 
and Belle, and the comparison between BESIII and Belle, where the Belle measurement has the $\eta\to\GG$ and $\eta\to
\pp\piz$ modes being combined.} 
\label{comparison}
\end{figure}

\item We show the comparisons among the latest results from Belle and the previous measurements from BaBar, 
BESIII and CMD-3 in Fig.~\ref{comparison}. The comparisons show good agreements in the four experiments.

\item In BaBar's measurements, the $\sigma(\EE\to\eta\phi)$ measured in $\eta\to\ppp$ mode is a bit lower than the 
one measured in the $\eta\to\GG$ mode, but both have similar lineshape, including a small bump around $2.13~\gev$. 
The expected $\jpsi$ signal according to the world average value of $\BR(\jpsi\to\eta\phi)$~\cite{PDG} is not clear 
in BaBar's measurements.

\item With a much large data sample, Belle's measurement is about twice as accurate as that measured in the BaBar
experiment. There are clear $\jpsi$ signals in both the $\eta\to\GG$ and $\ppp$ modes, while enhancement around 
neither $2.13~\gev$ nor $2.17~\gev$ is seen.

\item BESIII reported the Born cross section of $\EE\to\eta\phi$. We calculate the dressed cross section of $\EE \to 
\eta\phi$ with the vacuum polarization and the Born cross from Ref.~\cite{etaphi_bes3}, as shown in 
Fig.~\ref{xs_four_exps}(f).

\item The $50~\mev$ interval in BESIII data sample is a disadvantage in determining the lineshape of a structure with 
a width of tens $\mev$. Meanwhile, in the determination of $\phiy$, BESIII relied on $\sigma(\EE\to\eta\phi)$ below 
$2~\gev$ measured by the BaBar experiment, where large contribution from the $\phit$ signal dominates.

\item $\sigma(\EE\to\eta\phi)$ measured by CMD-3 are below $2~\gev$, with a precision similar to Belle's measurement.

\item Clear $\phit$ signals are observed in the BaBar, Belle, and CMD-3 measurements.

\eitm

The measurements of the dressed cross section of $\EE\to\eta\phi$ from the four experiments are consistent with each 
other. Therefore, we combine these measurements to get a best precision of $\sigma(\EE\to\eta\phi)$. The precise 
$\sigma(\EE\to\eta\phi)$ is helpful for studying the anomalous magnetic moment of muon~\cite{etaphi_cmd3}, and may 
show hints of $\phiy$ or $X(1750)$. The calculation for the combination is by
\beqar\label{eq_1}
\bar{x} & = &\frac{\sum_{i}{x_i/\Delta x^2_i}}{\sum_{i}{1/\Delta x^2_i}} , \\
\bar{\sigma} & = &\frac{\sum_{i}{\sigma_i/\Delta \sigma^2_i}}{\sum_{i}{1/\Delta \sigma^2_i}}, \\
(\Delta \sigma)^2 & = &\frac{1}{\sum_i {1/\Delta \sigma^2_i}}\label{eq_1c},
\eeqar
where $\sigma_i$ is the value of $i$th ($i=1,~2,~3,~4,~5,~6$) experimental measurement of cross section at energy 
point $x_i$ ($\sqrt{s_i}$) illustrated in Fig.~\ref{xs_four_exps}, $\Delta \sigma_i$ and $\Delta x_i$ are their 
related uncertainties. The average of $x_i$ takes into account the difference of $\sqrt{s}$ in the data taking 
in the BESIII or the CMD-3 experiment and the average $\sqrt{s}$ reported by BaBar and Belle using ISR technology. 
The uncertainties of $\sqrt{s}$ in BESIII and CMD-3 experiments are of $1~\mev$ level, and the two experiments have 
no overlap in $\sqrt{s}$ region. We take half of the $\sqrt{s}$ bin width in BaBar and Belle measurements as the 
uncertainty ($\Delta x_i$). However, there are correlations between the measurements, such as the branching 
fraction of $\phi$ or $\eta$ decay. We revisit the estimation of the Eq.~(\ref{eq_1c}) according to 
Ref.~\cite{Schmelling}. We construct the matrices of the statistics uncertainties $C^{stat}$ and the 
uncorrelated systematic uncertainties $C^{uncor\_syst} $ by 
\begin{equation}
C^{stat/uncor\_syst} =
\begin{pmatrix}
S_1\cdot \sigma_1^2	&	0	&	0	&	\cdots	&	0	\\
0	&	S_2\cdot\sigma_2^2	&	0	&	\cdots	&	0	\\
0	&	0	&	S_3\cdot\sigma_3^2	&	\cdots	&	0	\\
\vdots	&	\vdots	&	\vdots	&	S_i\cdot\sigma_i^2	&	\vdots	\\
0	&	0	&	0	&	\cdots	&	S_6\cdot\sigma_6^2	\\
\end{pmatrix}
\end{equation}
with $S_i = (\delta_i^{stat})^2$ or $(\delta_i^{uncor\_syst})^{2}$, where $\delta_i^{stat}$ and 
$\delta_i^{uncor\_syst}$ are the statistical and uncorrelated systematic relative uncertainties of $\sigma_i$. We 
then construct the matrix of the correlated systematic uncertainties $C^{cor\_syst}$ by
\begin{equation}
C^{cor\_syst} =
\begin{pmatrix}
a_{11}	&	a_{12}	&	\cdots	&	a_{1j}	\\
a_{12}	&	a_{22}	&	\cdots	&	a_{2j}	\\
\vdots	&	\vdots	&	\ddots	&	\vdots	\\
a_{i1}	&	a_{i2}	&	\cdots	&	a_{ij}	\\
\end{pmatrix}
\end{equation}
where $a_{ij} \equiv \delta_{i}^{cor\_syst}\cdot \delta_j^{cor\_syst}\cdot \sigma_{i}\cdot \sigma_{j}$. We get the 
effective global covariance matrix 
\begin{equation}\label{global_cov}
	\begin{aligned}
		C = C^{stat} + C^{uncor\_sys} + C^{cor\_syst}
	\end{aligned}
\end{equation}
According to Ref~\cite{Schmelling}, we calculate the error of $\bar{\sigma}$ by
\begin{equation}
	\begin{aligned}
	(\Delta \sigma)^{2} = (\sum_{ij} (C^{-1})_{ij})^{-1}.
	\end{aligned}
\end{equation}

\begin{figure}[htbp]
\psfig{file=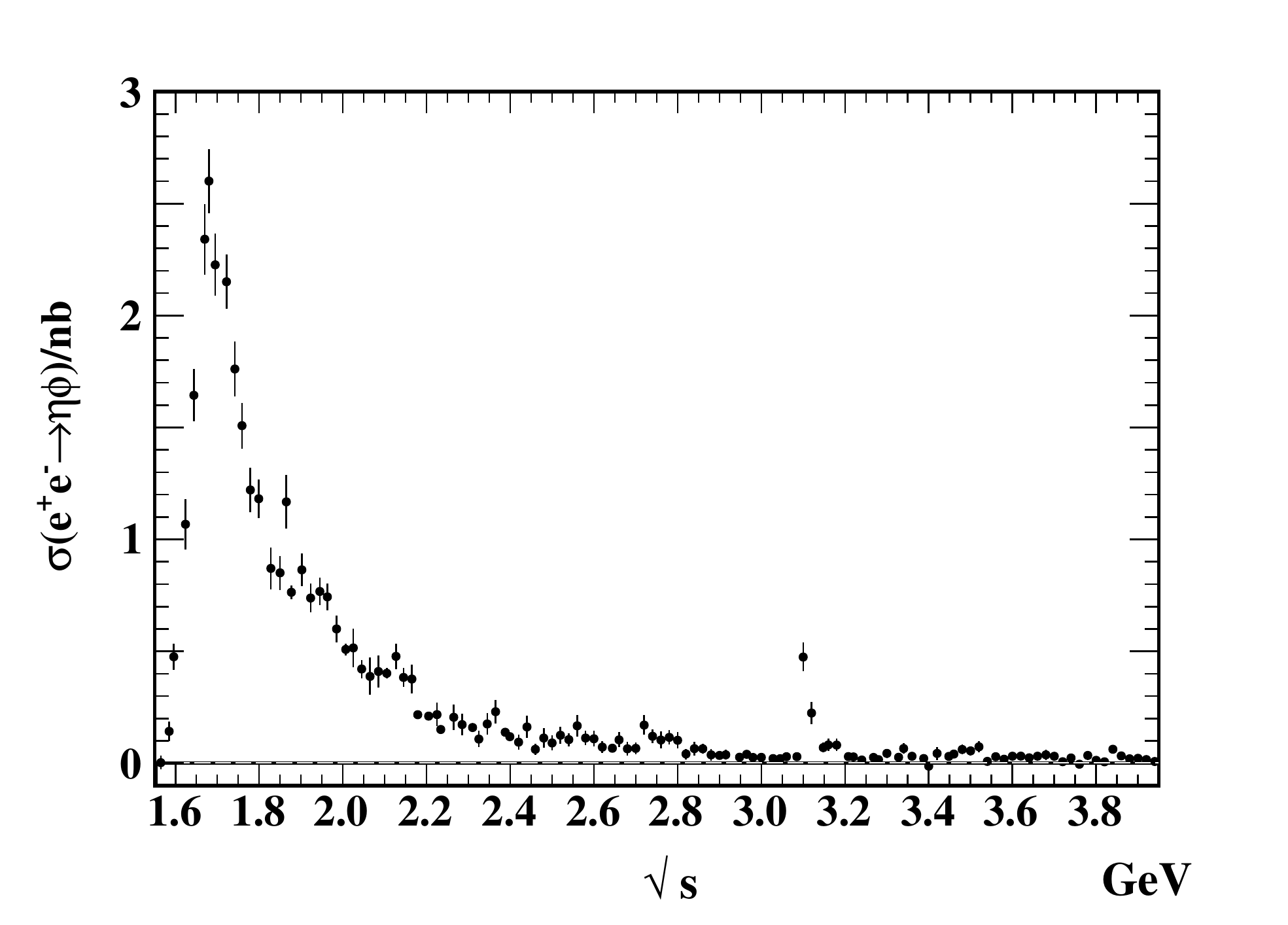,width=0.6\textwidth}
\caption{The cross section of $\EE \to \eta\phi$ from the combination of the measurements by the BaBar, Belle, BESIII 
and CMD-3 experiments.}
\label{cross}
\end{figure}

We show the results from combination in Fig.~\ref{cross} and Table~\ref{tab_xs_combine}.

\begin{table}[htbp]
\caption{The cross section of $\EE\to\eta\phi$ versus $\sqrt{s}$ calculated with the measurements from the BaBar, 
Belle, BESIII and CMD-3 experiments. The first errors are statistical and the second ones are systematic. }
\label{tab_xs_combine}
%\begin{center}
%\begin{small}
%\resizebox{\textwidth}{11cm}{
\begin{tabular}{c c | c c | c c | c c}
\hline\hline
$\sqrt{s}$ & $\sigma(\EE\to\eta\phi)$  & $\sqrt{s}$ & $\sigma(\EE\to\eta\phi)$ & $\sqrt{s}$ & $\sigma(\EE\to\eta\phi)$ & 
 $\sqrt{s}$ & $\sigma(\EE\to\eta\phi)$  \\\hline
1.565 &	$3\pm 30\pm 1$     & 2.165 & $376\pm 65\pm 22$  & 2.76 & $104\pm 38\pm 5$ & 3.36 & $31\pm 18\pm 2$  \\
1.585 & $143\pm 44\pm 4$   & 2.179 & $217\pm 17\pm 10$  & 2.78 & $116\pm 31\pm 7$ & 3.388 & $21\pm 15\pm 2$  \\
1.596 & $476\pm 60\pm 11$   & 2.205 & $211\pm 15\pm 8$  & 2.80 & $103\pm 36\pm 5$ & 3.40 & $-13\pm 8\pm 1$  \\
1.624 & $1068\pm 112\pm 36$ & 2.225 & $218\pm 53\pm 8$  & 2.82 & $42\pm 23\pm 1$  & 3.42 & $44\pm 29\pm 3$  \\
1.644 & $1644\pm 117\pm 53$ & 2.234 & $15\pm 13\pm 7$   & 2.84 & $65\pm 30\pm 3$  & 3.448 & $31\pm 12\pm 1$  \\
1.67  & $2341\pm 157\pm 90$ & 2.265 & $206\pm 56\pm 11$  & 2.86 & $65\pm 22\pm 3$  & 3.46 & $41\pm 19\pm 2$  \\
1.68  & $2600\pm 144\pm 76$ & 2.285 & $173\pm 48\pm 7$  & 2.88 & $38\pm 26\pm 2$  & 3.48 & $62\pm 22\pm 3$  \\
1.695 & $2226\pm 139\pm 62$ & 2.31 & $160\pm 11\pm 6$  & 2.90 & $35\pm 2\pm 2$   & 3.50 & $55\pm 20\pm 3$  \\
1.722 & $2151\pm 121\pm 59$ & 2.325 & $108\pm 37\pm 5$ & 2.915 & $38\pm 22\pm 3$  & 3.52 & $74\pm 24\pm 4$  \\
1.742 & $1761\pm 124\pm 48$ & 2.345 & $176\pm 48\pm 9 $ & 2.948 & $27\pm 5\pm 1$   & 3.54 & $8\pm 12\pm 7$   \\
1.759 & $1508\pm 102\pm 40$ & 2.365 & $230\pm 54 \pm 10$ & 2.965 & $40\pm 16\pm 4$  & 3.56 & $29\pm 14\pm 1$  \\
1.779 & $1221\pm 100\pm 34$ & 2.388 & $139\pm 10 \pm 5$ & 2.98 & $26\pm 5\pm 1$   & 3.58 & $18\pm 13\pm 3$  \\
1.799 & $1182\pm 86\pm 29$  & 2.399 & $119\pm 5\pm 5$   & 3.00 & $26\pm 5\pm 1$   & 3.60 & $32\pm 14\pm 1$  \\
1.828 & $870\pm 94\pm 24$   & 2.42 & $94\pm 33\pm 7$   & 3.028 & $21\pm 4\pm 2$   & 3.62 & $32\pm 19\pm 1$  \\
1.85  & $851\pm 76\pm 24$   & 2.44 & $162\pm 49\pm 8$  & 3.045 & $21\pm 4\pm 1$   & 3.64 & $25\pm 18\pm 1$  \\
1.865 & $1168\pm 120\pm 35$ & 2.46 & $62\pm 26\pm 8 $  & 3.06 & $29\pm 5\pm 1$   & 3.66 &  $32\pm 21\pm 2$ \\
1.877 & $764\pm 32\pm 12$   & 2.48 & $113\pm 44 \pm 6$ & 3.085 & $29\pm 2\pm 1$   & 3.68 & $38\pm 23\pm 2$  \\
1.902 & $864\pm 73\pm 22$   & 2.50 & $91\pm 33 \pm 2 $  & 3.10 & $475\pm 64\pm 24$ & 3.70 & $31\pm 18\pm 2$  \\
1.923 & $738\pm 63\pm 19$   & 2.52 & $125\pm 39 \pm 4$ & 3.12 & $224\pm 50\pm 10$ & 3.72 & $7\pm 9\pm 1$    \\
1.945 & $768\pm 61\pm 21$   & 2.54 & $106\pm 29 \pm 6$ & 3.148 & $70\pm 22\pm 3$  & 3.74 & $24\pm 12\pm 1$  \\
1.963 & $743\pm 60\pm 19$   & 2.56 & $167\pm 49 \pm 9$ & 3.16 & $82\pm 28\pm 4$  & 3.76 & $-4\pm 6\pm 1$   \\
1.985 & $600\pm 60\pm 16$   & 2.58 & $113\pm 32\pm 5 $ & 3.18 & $81\pm 28\pm 4$  & 3.78 & $35\pm 21\pm 3$  \\
2.007 & $509\pm 26\pm 12$   & 2.60 & $110\pm 36\pm 5 $ & 3.208 & $30\pm 13\pm 2$  & 3.80 & $13\pm 10\pm 1$  \\
2.025 & $515\pm 86\pm 26$   & 2.62 & $72\pm 26\pm 2 $  & 3.22 & $28\pm 16\pm 2$  & 3.82 & $6\pm 9\pm 1$    \\
2.045 & $421\pm 41\pm 15$   & 2.644 & $67\pm 5\pm 3 $   & 3.24 & $15\pm 16\pm 3$  & 3.84 & $63\pm 20\pm 3$  \\
2.065 & $388\pm 83\pm 29$   & 2.66 & $105 \pm 34\pm 5$ & 3.268 & $26\pm 12\pm 1$  & 3.86 & $33\pm 20\pm 2$  \\
2.085 & $410\pm 71\pm 20$   & 2.68 & $65\pm 30\pm 3$   & 3.28 & $16\pm 18\pm 1$  & 3.88 & $20\pm 11\pm 1$  \\
2.105 & $402\pm 24\pm 14$   & 2.70 & $67\pm 27\pm 3$   & 3.30 & $44\pm 20\pm 2$  & 3.90 & $22\pm 11\pm 1$  \\
2.127 & $478\pm 57\pm 12$   & 2.72 & $171\pm 44\pm 8$  & 3.328 & $27\pm 15\pm 2$  & 3.92 & $17\pm 10\pm 1$  \\
2.145 & $384\pm 42\pm 12$   & 2.74 & $121\pm 32\pm 7$  & 3.34 & $66\pm 24\pm 3$  & 3.94 & $8\pm 13\pm 1$   \\
\hline\hline
\end{tabular}
%}
%\end{center}
%\end{small}
\end{table}

\section{Parametrization of $\sigma(\EE\to\eta\phi)$}
\label{sec_metaphi}

There could be $\phit$, $X(1750)$ and $\phiy$ in the $\EE\to\eta\phi$ process. We perform combined fits to the 
$\sigma(\EE\to\eta\phi)$ measured by BaBar, Belle, BESIII and CMD-3 experiments, and shown in 
Fig.~\ref{xs_four_exps}. The fit range is from the threshold to $2.85~\gevcs$. Assuming there are $\phit$, $X(1750)$, 
$\phiy$ components and non-resonant contribution in the $\eta\phi$ final state, we take the parametrization of 
$\sigma(\EE\to\eta\phi)$ similar to that used in BaBar's analysis~\cite{etaphi_babar}:
\beq\label{eq_2}
\sigma(\EE\to\eta\phi)(\sqrt{s}) =
12\pi\mathcal{P}_{\eta\phi}(\sqrt{s})|A_{\eta\phi}^{n.r.}(\sqrt{s})+ A_{\eta\phi}^{\phit}(\sqrt{s}) + 
A_{\eta\phi}^{X(1750)}(\sqrt{s}) +
A_{\eta\phi}^{\phiy}(\sqrt{s}) |^2,
\eeq
where $\mathcal{P}_{\eta\phi}$ is the phase space of the $\eta\phi$ final state, the non-resonant amplitude takes the 
form $A_{\eta\phi}^{n.r.}(\sqrt{s}) = a_{0}/s^{a_{1}}$, and $A_{\eta\phi}^{\phit}$, $A_{\eta\phi}^{X(1750)}$ and 
$A_{\eta\phi}^{\phiy}$ are the amplitudes of the $\phit$, $X(1750)$ and $\phiy$, respectively.

For $A_{\eta\phi}^{\phit}$ and $A_{\eta\phi}^{X(1750)}$, we describe the form with a Breit-Wigner (BW) function
\beq\label{eq_3}
A^{\eta\phi}_X(\sqrt{s}) = \sqrt{\BR^{\eta\phi}_{X}\Gamma^{\EE}_{X}}\cdot \frac{ \sqrt{\Gamma_X/
\mathcal{P}_{\eta\phi}(M_X)}\cdot e^{i\theta_{X}}}{M_{X}^2 - s - i\sqrt{s}\Gamma_{X}(\sqrt{s})},
\eeq
where $X$ is $\phit$ or $X(1750)$, the resonant parameters $M_X$, $\Gamma_X$ and $\Gamma_X^{\EE}$ are the mass, the
total width and the partial width to $\EE$, respectively. $\BR_X^{\eta\phi}$ is the branching fraction of $X \to
\eta\phi$ decay and $\theta_X$ is the relative phase.

BaBar's measurement~\cite{etaphi_babar} shows that $\kkt$ and $\eta\phi$ are two major decays of $\phit$ and 
$\BR^{\kkt}_{\phit} \approx 3 \times \BR^{\eta\phi}_{\phit}$, where $\BR^{\kkt}_{\phit}$ is the branching 
fraction of $\phit \to \kkt$ decay. We also take the form as in Ref~\cite{etaphi_babar}:
\beqar\label{eq_4}
\Gamma_{\phit}(\sqrt{s}) & = &
\Gamma_{\phit}\cdot [\frac{\mathcal{P}_{KK^{*}(892)}(\sqrt{s})}{\mathcal{P}_{\kkt}(M_{\phit})}\BR^{\kkt}_{\phit} +
\frac{\mathcal{P}_{\eta\phi}(\sqrt{s})}{\mathcal{P}_{\eta\phi}(M_{\phit})}\BR^{\eta\phi}_{ \phit} \nonumber \\
 & & + (1-\BR^{\eta\phi}_{\phit}-\BR^{\kkt}_{\phit})].
\eeqar
Here, $\mathcal{P}_{\kkt}$ is the phase space of the $\phit\to \kkt$ decay. The other decays of $\phit$ are 
neglected, and their phase space dependence correspondingly are ignored. Since both the $\kkt$ and $\eta\phi$ final 
states contain a vector meson ($V$) and a pseudoscalar meson ($P$), the phase space takes the form
\beq\label{eq_5}
\mathcal{P}_{VP}(\sqrt{s})=[\frac{(s + M^2_V - M^2_P)^2 - 4 s M^2_V} s]^{3/2}.
\eeq

We take the form of $X(1750)$ as in BESIII's measurement~\cite{kketa}:
\beq\label{eq_6}
\Gamma_{X(1750)}(\sqrt{s}) = \Gamma_{X(1750)}\cdot \frac{M_{X(1750)}^2}{s} 
\cdot [\frac{p(\sqrt{s})}{p(M_{X(1750)})}]^{2l+1},
\eeq
where $p(\sqrt{s})$ [$p(M_{X(1750)})$] is the momentum of a daughter particle in the rest frame of the resonance with 
energy $\sqrt{s}$ (mass $M_{X(1750)}$), and $l$ is the orbital angular momentum of the daughter particle.

We describe the amplitude of $\phiy\to\eta\phi$ decay as in Ref.~\cite{etaphi_belle}:
\beq\label{eq_7}
A_{\eta\phi}^{\phiy}(s) =
\sqrt{\BR^{\eta\phi}_{\phiy}\Gamma^{\EE}_{\phiy}}\cdot \frac{\sqrt{\Gamma_{\phiy}/\mathcal{P}_{\eta\phi}[M_{\phiy}]}
\cdot e^{i\theta_{\phiy}}}{M^2_{\phiy} - s - i\sqrt{s}\Gamma_{\phiy}} \cdot\frac{B(p)}{B(p')},
\eeq
where $M_{\phiy}$ and $\Gamma_{\phiy}$ are the mass and width of $\phiy$, $B(p)$ is the $P$-wave Blatt-Weisskopf form
factor and $p$ ($p'$) is the breakup momentum corresponding to the $\sqrt{s}$ (mass $M_{\phiy}$).

\section{Fit results for $\phit$, $\phiy$ and $X(1750)$}

We perform several combined fits to the $\sigma(\EE\to\eta\phi)$ measured by BaBar, Belle, BESIII and CMD-3. They are 
fits with 1) only $\phit$; 2) $\phit$ and non-resonant component; 3) $\phit$, $\phiy$ and non-resonant component; 
4) $\phit$, $X(1750)$, $\phiy$ and non-resonant component. The input data are the $\sigma(\EE\to\eta\phi)$ and the 
related uncertainties shown in Fig.~\ref{xs_four_exps}. According to the fit results, which will be described below, 
we get the nominal fit results from the third case. 

The input data of the combined fits are the values of $\sigma(\EE\to\eta\phi)$ measured by BaBar, Belle, BESIII 
and CMD-3 experiments, and a least $\chi^2$ method with MINUIT~\cite{minuit} is used. 
According to 
Ref.~\cite{Schmelling}, we define the $\chi^2_k$ of the $k$th energy point as 
\beq\label{eq_8}
\chi^2_k = \sum_{i,j}(\Delta\sigma_{ik})(C^{-1})_{ij}(\Delta \sigma_{jk}), 
\eeq
%\beq\label{eq_8}
%\chi^{2}= \sum_{i}\sum_{j} \sum_{k} \Delta \sigma_i\cdot \xi_{ijk}\cdot \Delta \sigma_j\cdot \xi_{jik},
%\eeq
where $\Delta \sigma_{ik}$ ($\Delta \sigma_{jk}$) is the difference between the measured value from the $i$th ($j$th) 
data sample and the fitted value of $\sigma(\EE\to\eta\phi)$, and the effective global covariance matrix $C$ is 
described in Eq.~(\ref{global_cov}). The total $\chi^2$ is the sum of $\chi^2_k$ over all energy points.

In the measurement from one experiment, there could be correlation of the two modes of $\eta$ decays in one 
$\sqrt{s}$ bin, or correlation of all $\sqrt{s}$ bins. For the first correlation, we calculate  
\beq
\chi^{\prime 2}_i = (C_{ii}^{\prime})\cdot (\sigma^{\GG,i}_{measured} - \sigma^i_{fit})\cdot 
(\sigma^{\pp\piz,i}_{measured} - \sigma^i_{fit})
\eeq
for the $i$th $\sqrt{s}$ bin. Here, we also use the relative uncertainty $(\delta_i^{cor\_syst})^2$ between the two 
modes of $\eta$ decays to calculate the elements $C_{ii}^{\prime}$ of the correlation matrix. We get $\chi^{\prime 2} 
= \sum_i \chi^{\prime 2}_i$ for the sum of all the $\sqrt{s}$ bins in one experiment. Similarly, we calculate the 
$\chi^{\prime \prime 2}$ for the second correlation to be
\beq
\chi^{\prime \prime 2} = \sum_{i,j} (C_{ij}^{\prime \prime}) \cdot (\sigma_{measured}^{i} - \sigma_{fit}^i)\cdot 
(\sigma_{measured}^{j} - \sigma_{fit}^j),~ i\ne j.
\eeq
Here, the matrix element $C_{ij}^{\prime \prime} = \delta_{i}^{cor\_syst}\cdot \delta_{j}^{cor\_syst}$ is for the 
correlation between the $i$th and the $j$th $\sqrt{s}$ bins. Need to mention that $\delta_{i}^{cor\_syst}$ refers to 
different correlated systematic uncertainties in the calculations of $\chi^{\prime 2}$ and $\chi^{\prime \prime 2}$.

We add $\chi^{\prime 2}$ and $\chi^{\prime \prime 2}$ to the total $\chi^2$ for the constraints due to the two kinds 
correlation in the combined fits.

%\mkred{According to 
%Ref.~\cite{Schmelling}, we define the \mkred{$\chi^2_k$ of the $k$th energy point} as 
%\mkred{\beq\label{eq_8}
%\chi^2_k = \sum_{i,j}(\Delta\sigma_{ik})(C^{-1})_{ij}(\Delta \sigma_{jk}). ~\\
%\eeq}
%%\beq\label{eq_8}
%%\chi^{2}= \sum_{i}\sum_{j} \sum_{k} \Delta \sigma_i\cdot \xi_{ijk}\cdot \Delta \sigma_j\cdot \xi_{jik},
%%\eeq
%where $\Delta \sigma_{ik}$ ($\Delta \sigma_{jk}$) is the difference between the measured values and the 
%fitted values of $\sigma(\EE\to\eta\phi)$, and the effective global covariance matrix $C_{ij}$ is 
%described in Eq.~(\ref{global_cov}). The total $\chi^2$ is the sum of $\chi^2_k$ over all energy points.} 
%We treat all the statistical and the systematic uncertainties irrelevant, except the systematic uncertainty of 
%$\BR[\phi\to K^+K^-]$. When $i=j$, $\xi_{ijk}$ contains the statistical and uncorrelated systematic uncertainties. 
%When $i\ne j$, $\xi_{ijk} = \xi_{jik}$ is the common systematic uncertainty of $\sigma_i$ and $\sigma_j$ from source 
%$k$, and we choose the minimum value of the uncertainties at $i$-th and $j$-th energy points of the same source 
%since correlated relative uncertainty can not be larger than any measurements total relative uncertainty. 

\subsection{Fit with only $\phit$}

\begin{figure}[tbp]
\includegraphics[width=0.6\textwidth]{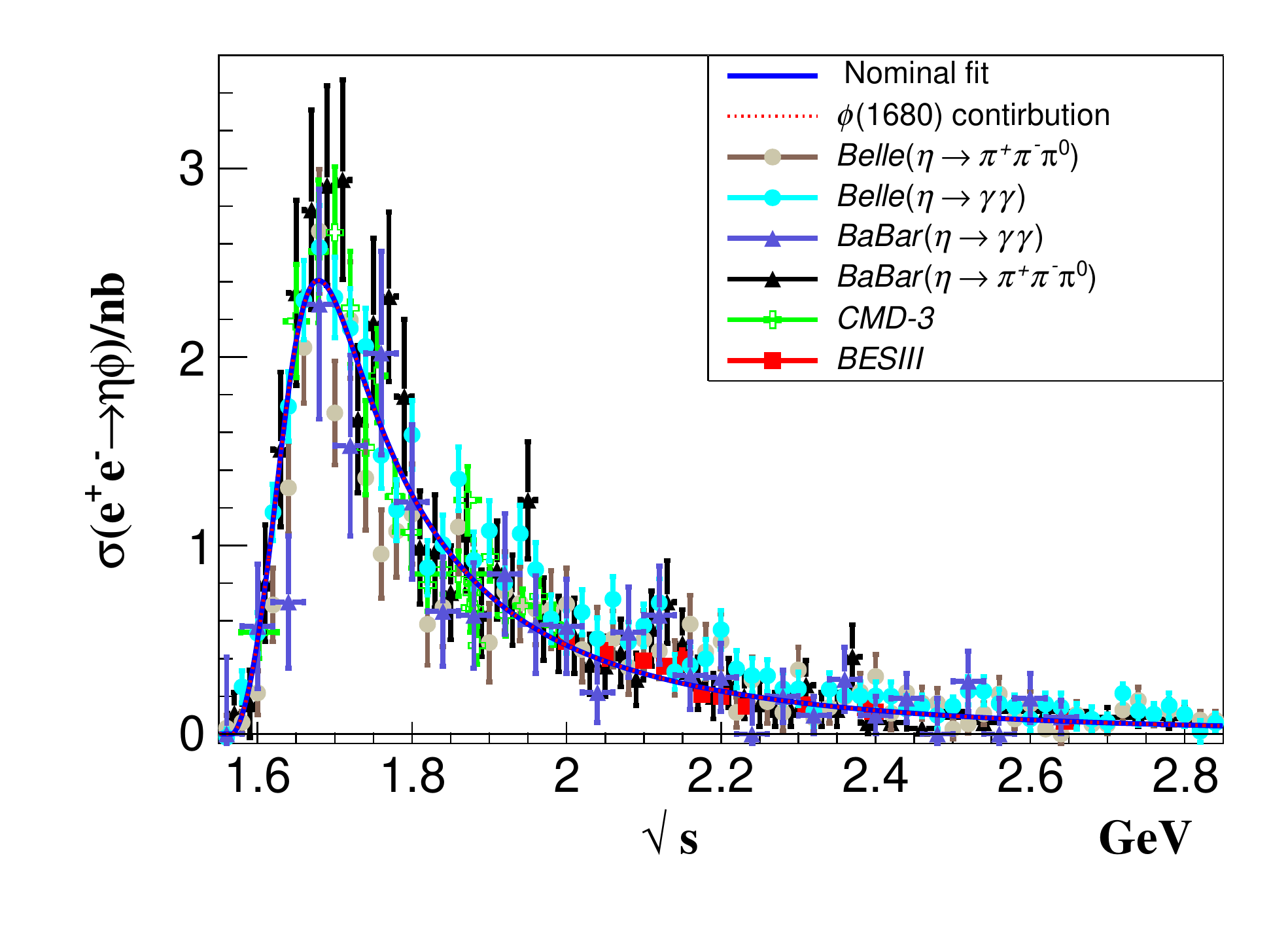}
\caption{Results of fitting to the $\sigma(\EE\to\eta\phi)$ measured by the BaBar, Belle, BESIII and CMD-3 
experiments with only $\phit$. The blue solid lines show the fit results, and the dashed red line shows the $\phit$ 
component.}
\label{onlyfit}
\end{figure}

Fitting to the $\sigma(\EE\to\eta\phi)$ measured by the four experiments with only the 
$A_{\eta\phi}^{\phit}(\sqrt{s})$ component in Eq.~\ref{eq_2}, we get reasonably good results with the quality of 
$\chi^2/ndf = 381/254$, as illustrated in Fig.~\ref{onlyfit}. Here $ndf$ is the number of all fitted data points 
minus the number of free parameters. We obtain the resonant parameters of $\phit$: $M_{\phit} = (1723 \pm 6)~\mevcs$, 
$\Gamma_{\phit} = (376 \pm 9)~\mev$ and $\BR_{\phit}^{\eta\phi} \Gamma^{\EE}_{\phit} = (197 \pm 5)~\ev$. The world 
average values of the mass and width of $\phit$ are $(1680\pm 20)~\mevcs$ and $(150\pm 50)~\mev$~\cite{PDG}, 
respectively. We can see that the mass and width from this fit are much different to the world average values, which 
is due to the absence of some components in our fit, such as the non-resonant contribution and the $\phiy$. We also 
notice that the world average value of the width has a large uncertainty. 

\subsection{Fit with $\phit$ and non-resonant component}

\begin{figure}[tbp]
\includegraphics[width=0.4\textwidth]{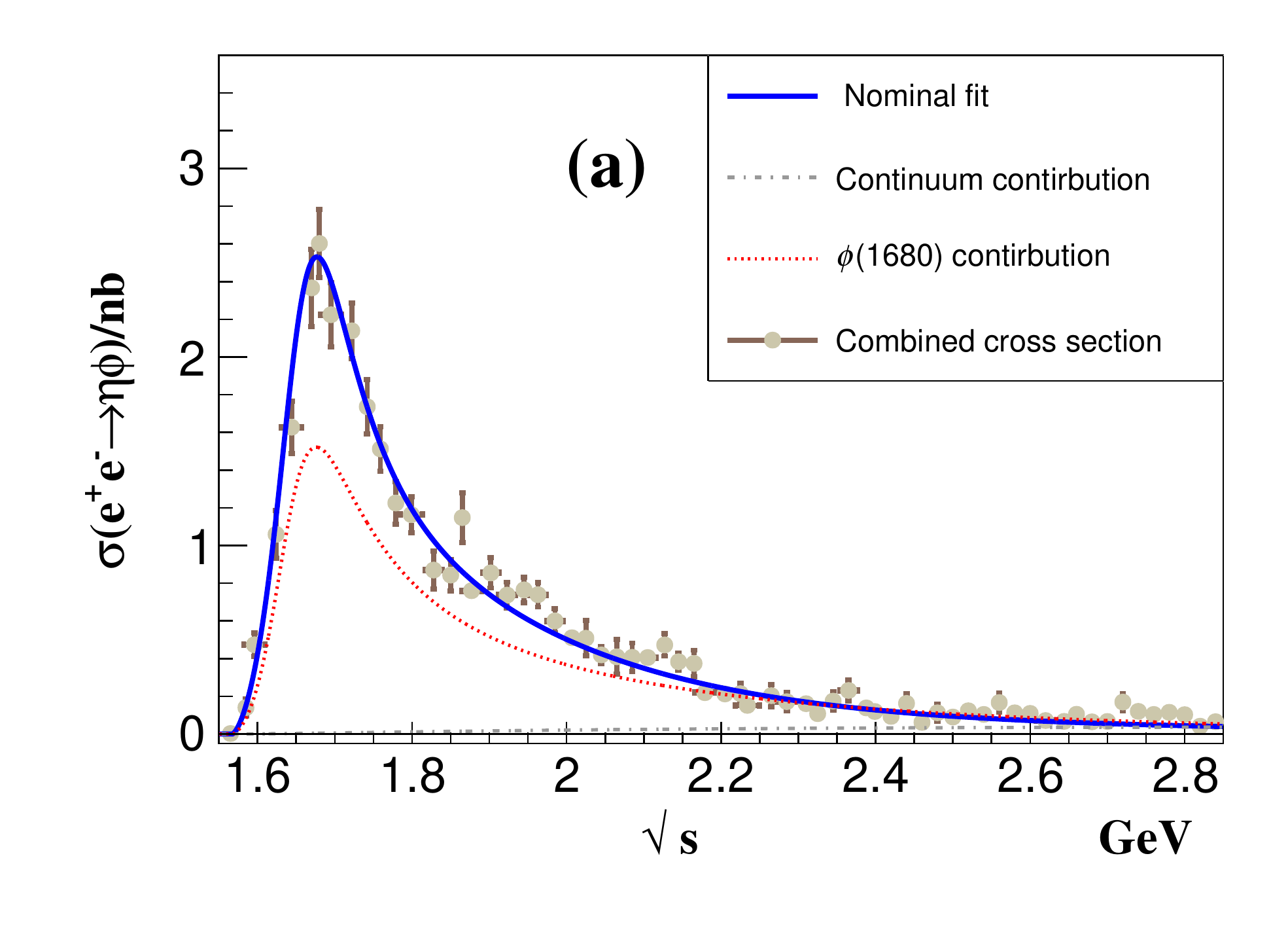}
\includegraphics[width=0.4\textwidth]{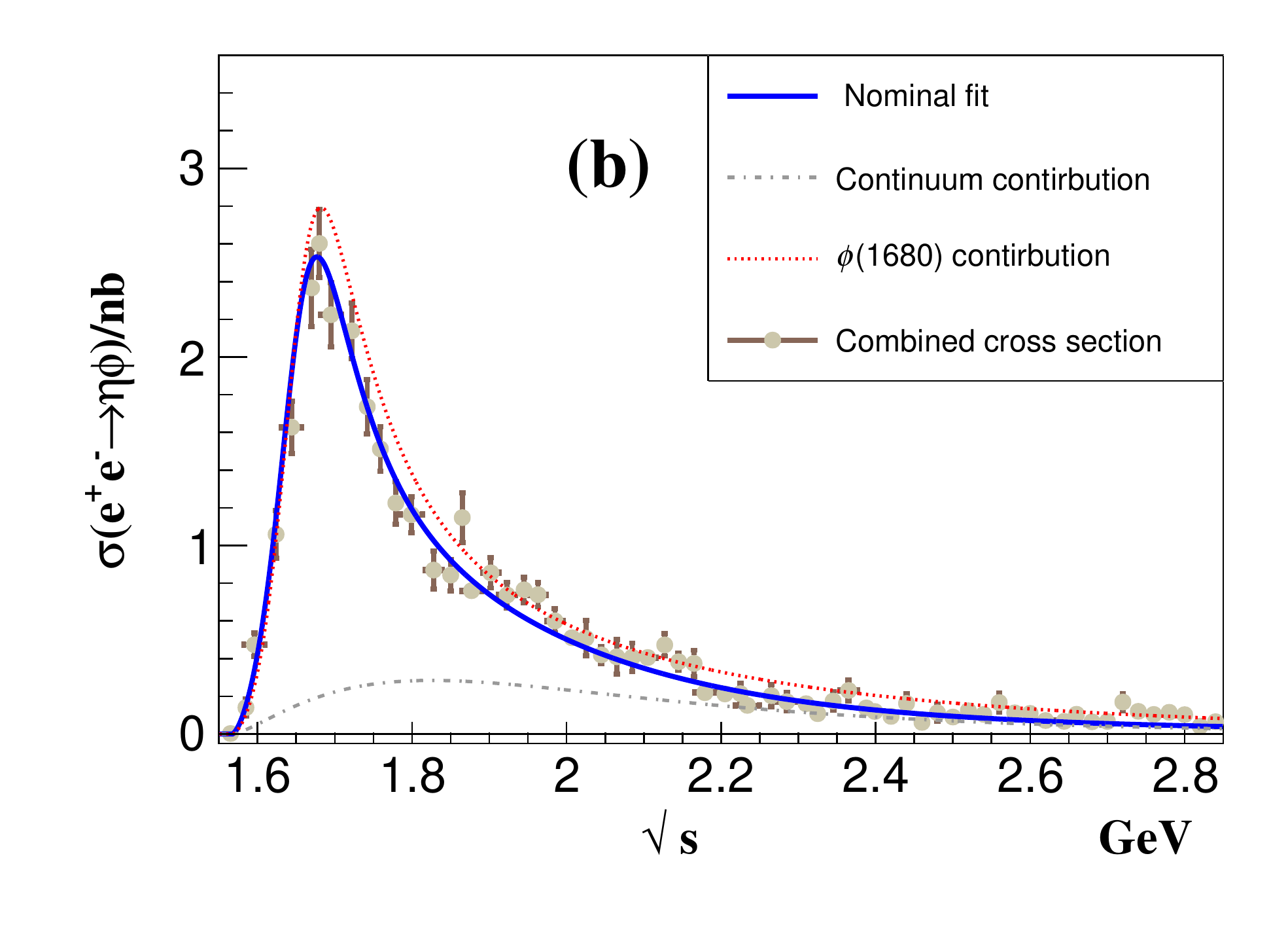}
\caption{Fitting to the $\sigma(\EE\to\eta\phi)$ measured by the BaBar, Belle, BESIII and CMD-3 experiments including 
the $\phit$ and the non-resonant contribution. The blue solid lines show the fit results, and the dashed red, green 
and gray lines show the $\phit$ and non-resonant components, respectively. The interference between non-resonant 
component and $\phit$ are not shown.}
\label{twofit}
\end{figure}

Fitting to the $\sigma(\EE\to\eta\phi)$ with only $A_{\eta\phi}^{\phit}(\sqrt{s})$ and the non-resonant contribution 
in Eq.~\ref{eq_2}, we get two solutions of equivalent quality with $\chi^2/ndf = 347/251$, as illustrated in 
Fig.~\ref{twofit} and Table~\ref{fit_results}. Here and hereinafter, we keep using all the measured data from the 
four experiments as input for the combined fits but show only the combined $\sigma(\EE\to \eta \phi)$ from 
Fig.~\ref{cross} to represent the data in the plots. We get the same resonant parameters $M_{\phit} = (1676 \pm 3)
~\mevcs$ and $\Gamma_{\phit} = (161^{+5}_{-4})~\mev$, while $\BR_{\phit}^{\eta\phi} \Gamma^{\EE}_{\phit} = (88 \pm 3)
~\ev$ or $(162^{+5}_{-3})~\ev$ from the two solutions. The two resonant parameters have good agreements to the world 
average values, and the precision is well improved. Meanwhile, the branching fraction of $\phit\to\eta\phi$ is 
$\BR_{\phit}^{\eta\phi} = (20^{+4}_{-3})\%$ or $(24\pm 3)\%$, which is close to the value $\sim 17\%$ that can be 
calculated according to BaBar's measurement~\cite{etaphi_babar_2}. We can see most of the measured $\sigma(\EE\to\eta
\phi)$ from the four experiments are above the fit curve around $2.17~\gev$ in Fig.~\ref{twofit}, which indicates the 
requirement of $\phiy$.

\subsection{Fit with $\phit$, $\phiy$ and non-resonant component}

\begin{figure}[tbp]
\includegraphics[width=0.4\textwidth]{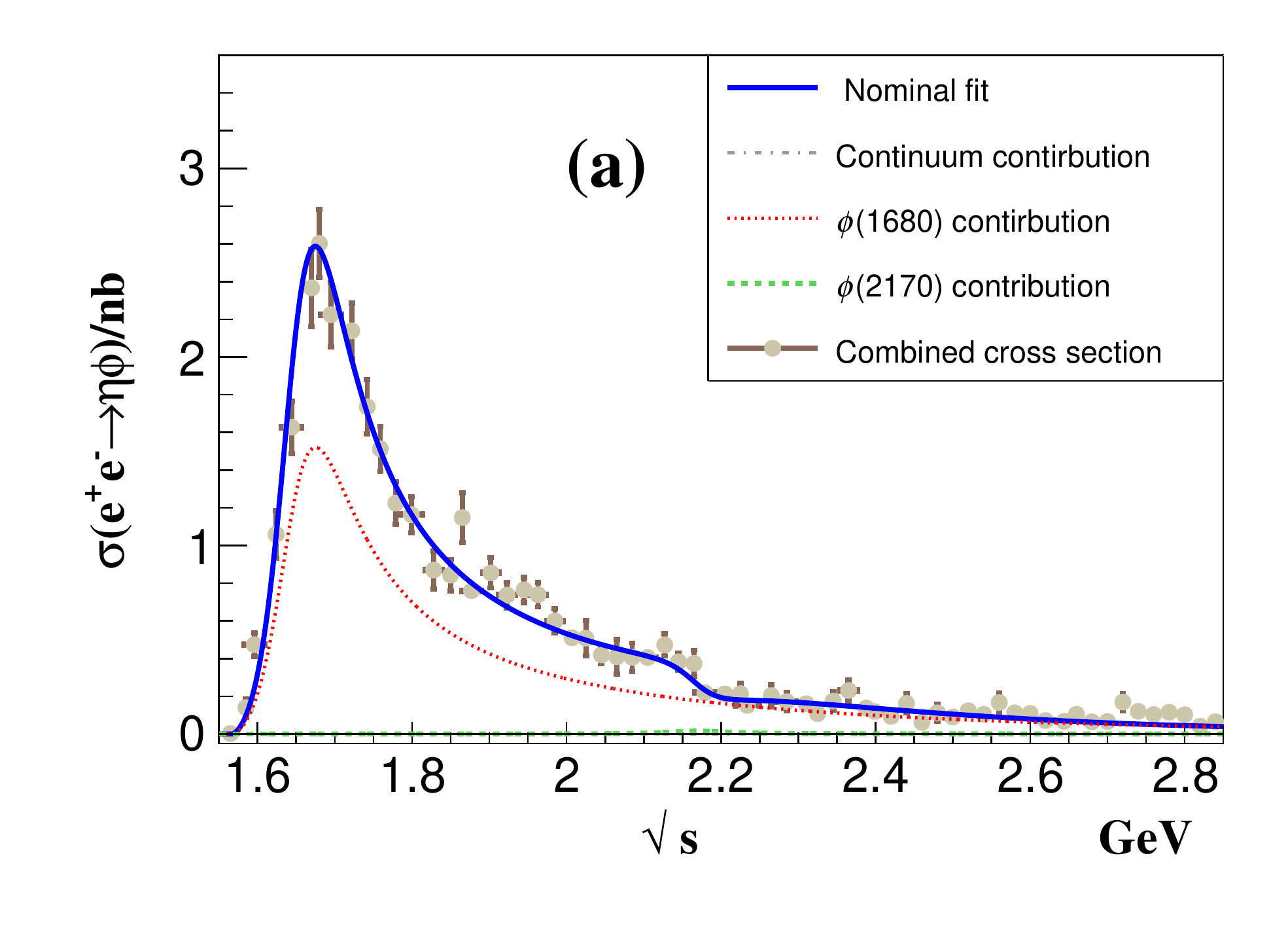}
\includegraphics[width=0.4\textwidth]{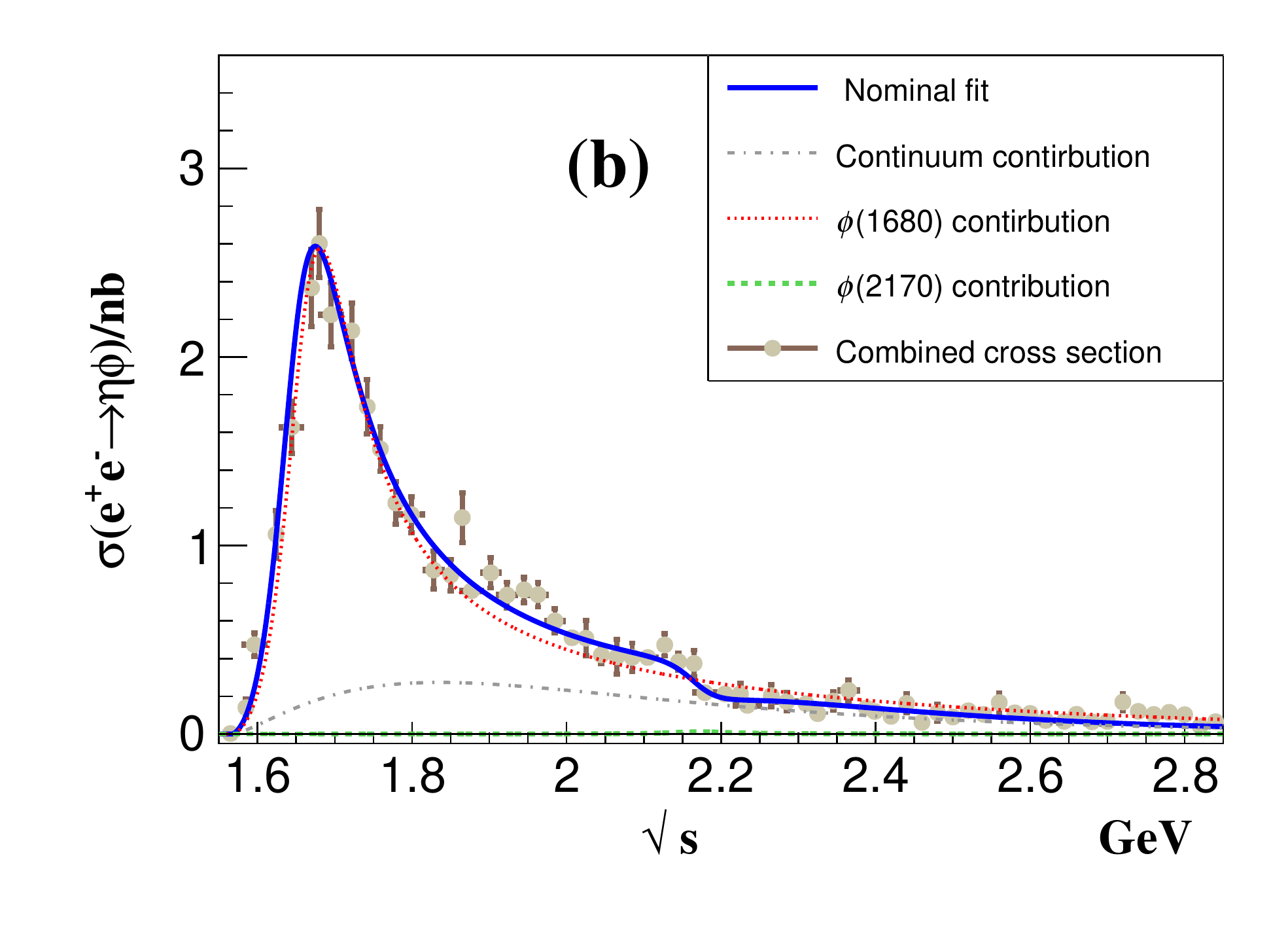}\\
\includegraphics[width=0.4\textwidth]{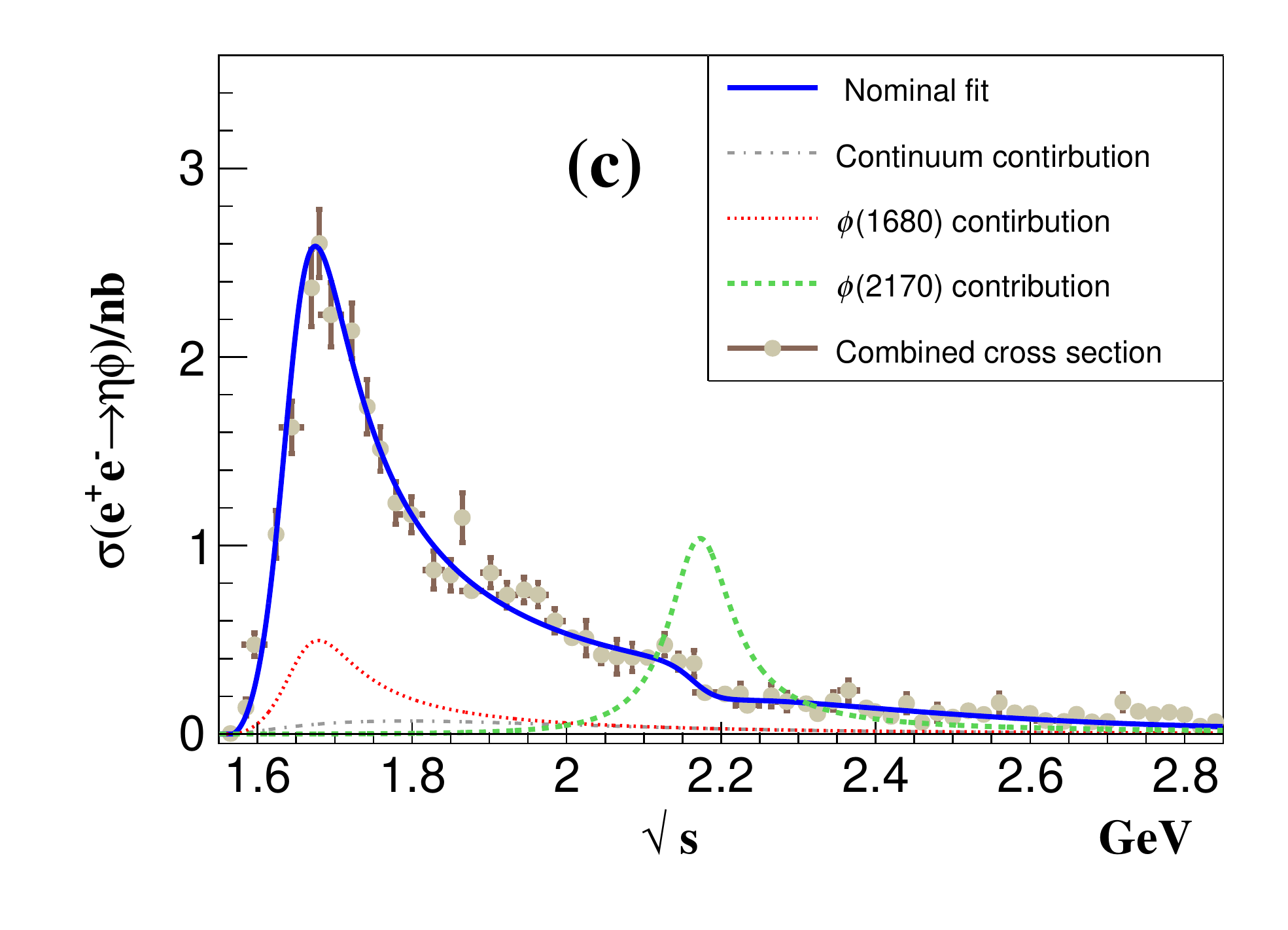}
\includegraphics[width=0.4\textwidth]{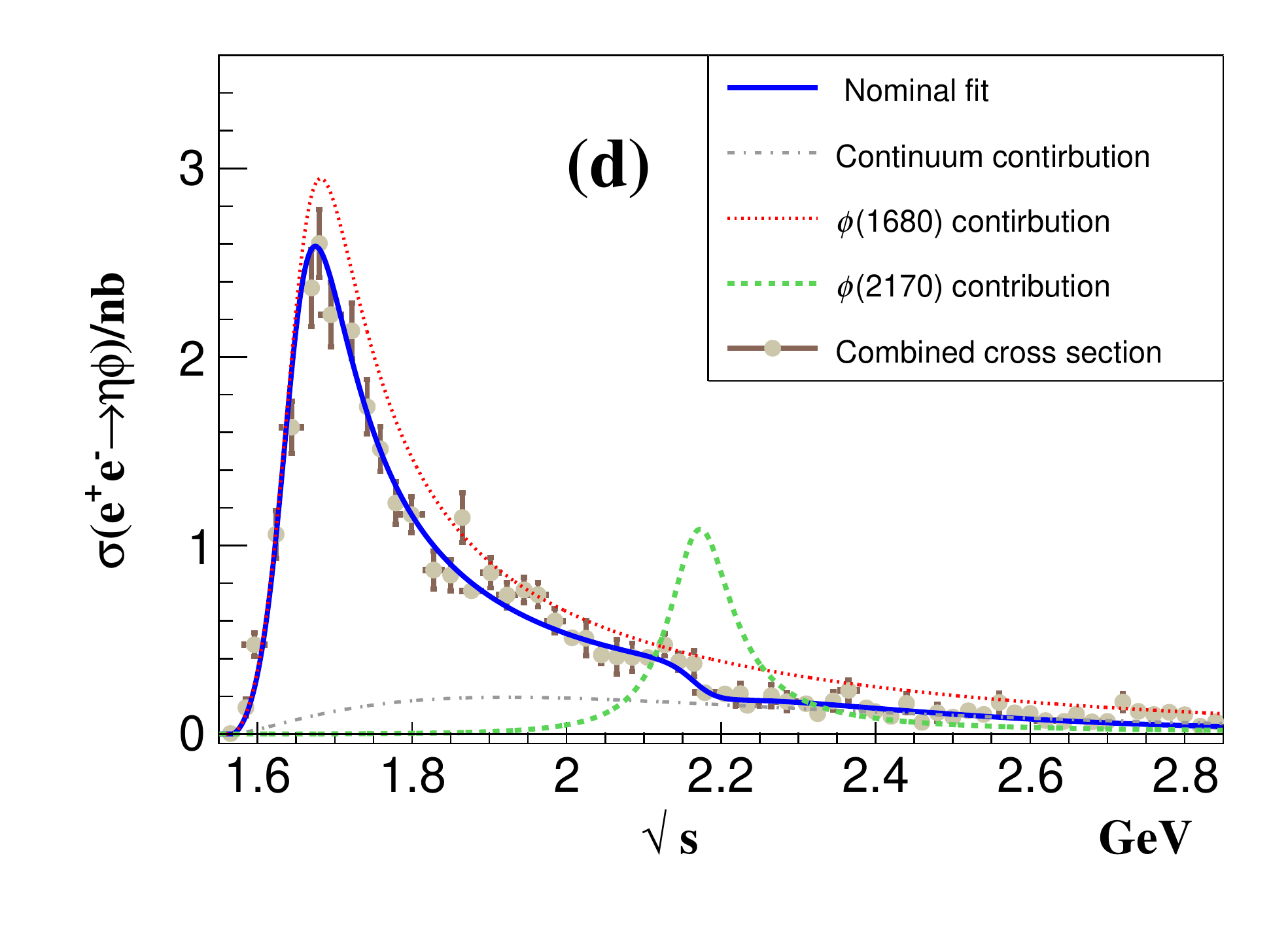}
\caption{Nominal fit to the $\sigma(\EE\to\eta\phi)$ measured by the BaBar, Belle, BESIII and CMD-3 experiments, 
including $\phit$, $\phiy$ and no-resonant components. The blue solid lines show the fit results, and the dashed red, 
green and gray lines show the $\phit$, $\phiy$ and non-resonant components, respectively. The interference among 
$\phit$, $\phiy$ and non-resonant components are not shown. }
\label{fit}
\end{figure}

With $A_{\eta\phi}^{\phit}$ and $A_{\eta\phi}^{\phiy}$  but no $A_{\eta\phi}^{X(1750)}$ in Eq.~\ref{eq_2}, we get 
four solutions of equivalent quality with $\chi^2/ndf = 284/247$ from the nominal combined fit, as illustrated in 
Fig.~\ref{fit} and Table~\ref{fit_results}. The four solutions have the same resonant parameters $M_{\phit}$, 
$\Gamma_{\phit}$, $M_{\phiy}$ and $\Gamma_{\phiy}$: $M_{\phit} = (1678^{+5}_{-3} \pm 7)~\mevcs$, $\Gamma_{\phit} = 
(156 \pm 5 \pm 9)~\mev$, $M_{\phiy} = (2169 \pm 5 \pm 6)~\mevcs$ and $\Gamma_{\phiy} = (96^{+17}_{-14} \pm 9)~\mev$. 
We can see that the mass and width of $\phiy$ are close to the world average values: $m_{\phiy} = (2162\pm 7)~\mevcs$ 
and $\Gamma_{\phiy} = (100^{+31}_{-23})~\mev$~\cite{PDG}. The four solutions show that $\BR_{\phit}^{\eta \phi} 
\Gamma^{\EE}_{\phit} = (79 \pm 4 \pm 16)$, $(127\pm 5 \pm 12)$, $(65^{+5}_{-4}\pm 13$ or $(215^{+8}_{-5}\pm 11)~\ev$, 
and $\BR_{\phiy}^{\eta\phi} \Gamma^{\EE}_{\phiy} = (0.56^{+0.03}_{-0.02} \pm 0.07)$, $(0.36 \pm 0.04 \pm 0.07)$, 
$(38\pm 1\pm 5)$ or $(41 \pm 2 \pm 6)~\ev$. The branching fraction of $\phit\to\eta\phi$ is $\BR_{\phit}^{\eta\phi} 
\approx 20\%$ with uncertainties less than 5\%. Comparing the change of $\Delta \chi^2 = 63$ and $\Delta ndf = 4$ 
between the fit with and without $\phiy$, we obtain the statistical significance of the $\phiy$ resonance to be 
$7.2\sigma$. By fixing the mass and width of $\phiy$ to the world average values~\cite{PDG}, we obtain the fit 
results listed in Table~\ref{fit_results1} with curves very similar to those in Fig.~\ref{fit}. We then estimate the 
statistical significance of the $\phiy$ to be $7.4\sigma$. We will describe the systematic uncertainties in the fit 
results in Sec.~\ref{sys_err}.

\begin{table}[htbp]
\begin{center}
\caption{Results of fitting to the $\sigma(\EE\to\eta\phi)$ measured by BaBar, Belle, BESIII and CMD-3 
experiments with non-resonant component, $\phit$ and $\phiy$, or without $\phiy$.}
\begin{tabular}{c|cccc | cc}\hline\hline
Parameters      &  \multicolumn{4}{c|}{with $\phiy$}  & \multicolumn{2}{c}{without $\phiy$}     \\\hline
                & Solution I   & Solution II   	& Solution III    & Solution IV  & Solution I & Solution II \\
$\chi^2/ndf$  & \multicolumn{4}{c|}{284/247}  & \multicolumn{2}{c}{347/251} \\
$a_{0}$ & $-0.12\pm 0.02$ & $4.4\pm 0.2$ & $1.1\pm 0.2$   & $-5.0\pm 0.2$ & $-0.9^{+0.1}_{-0.3}$ & $-5.0\pm 0.3$  \\
$a_{1}$ & $-4.8^{+0.2}_{-0.1}$ & $2.8\pm 0.1$ & $-3.1\pm 0.2$ & $2.6\pm 0.3$ & $1.0^{+0.4}_{-0.3}$ & $3.0\pm 0.1$  \\
$\BR^{\eta\phi}_{\phit)}\Gamma^{\EE}_{\phit} (\ev)$
              & $79\pm 4$ &  $127\pm 5$  & $65^{+5}_{-4}$ & $215^{+8}_{-5}$ & $88\pm 3$ & $162^{+5}_{-3}$   \\
$M_{\phit} (\mevcs)$  & \multicolumn{4}{c|}{$1678^{+5}_{-3}$}  & \multicolumn{2}{c}{$1676\pm3$} \\
$\Gamma_{\phit}(\mev)$ & \multicolumn{4}{c|}{$156\pm5$}  & \multicolumn{2}{c}{$161^{+5}_{-4}$} \\
$\BR^{\eta\phi}_{\phit}(\%)$
              & $19^{+2}_{-1}$ & $22\pm 2$ & $24^{+4}_{-3}$ & $19^{+2}_{-1}$ & $20^{+4}_{-3}$ & $24\pm 3$  \\
$\BR_{\phiy}^{\eta\phi}\Gamma^{\EE}_{\phiy} (\ev)$
              & $0.56^{+0.03}_{-0.02}$ & $0.36^{+0.05}_{-0.03}$ & $38\pm 1$   & $41\pm2$  & \multicolumn{2}{c}{---} \\
$M_{\phiy} (\mevcs)$    & \multicolumn{4}{c|}{$2169\pm5$}  & \multicolumn{2}{c}{---} \\
$\Gamma_{\phiy} (\mev)$ & \multicolumn{4}{c|}{$96^{+17}_{-14}$}   & \multicolumn{2}{c}{---} \\
$\theta_{\phit}(^\circ)$ & $-63\pm12$ & $-95\pm 9$ & $-88\pm6$ & $-122\pm7$ & $102^{+7}_{-6}$ & $-94^{+11}_{-6}$ \\
$\theta_{\phiy}(^\circ)$  & $81^{+14}_{-9}$  & $-77^{+10}_{-5}$ & $-159^{+19}_{-15}$ & $133^{+16}_{-13}$ & \multicolumn{2}{c}{---}  \\
\hline\hline
\end{tabular}
\label{fit_results}
\end{center}
\end{table}

\begin{table}[htbp]
\begin{center}
\caption{
Results of fitting to the $\sigma(\EE\to\eta\phi)$ measured by BaBar, Belle, BESIII and CMD-3 experiments with 
$\phit$, $\phiy$ and non-resonant component. The mass and width of $\phiy$ are fixed to the world average 
values~\cite{PDG}.}
\begin{tabular}{c|cccc} \hline \hline
Parameters      &  \multicolumn{4}{c}{with $\phiy$}       \\\hline
                & Solution I   & Solution II   & Solution III    & Solution IV    \\
$\chi^2/ndf$  & \multicolumn{4}{c}{288/249}   \\
$a_{0}$         & $-0.45\pm 0.05$ & $0.19\pm 0.02$   & $-4.1\pm 0.2$    & $-4.4\pm 0.5$   \\
$a_{1}$         & $-0.27^{+0.05}_{-0.04}$ & $2.8\pm 0.2$ & $-0.43^{+0.08}_{-0.05}$ & $2.6^{+0.5}_{-0.3}$    \\
$\BR^{\eta\phi}_{\phit}\Gamma^{\EE}_{\phit} (\ev)$
		    & $85\pm3$ & $123^{+6}_{-4}$ & $53\pm 6$ & $193^{+6}_{-5}$ \\
$M_{\phit} (\mevcs)$      & \multicolumn{4}{c}{$1677^{+5}_{-4}$}   \\
$\Gamma_{\phit}(\mev)$    & \multicolumn{4}{c}{$158^{+5}_{-4}$}   \\
$\BR^{\eta\phi}_{\phit}(\%)$
            & $18^{+5}_{-3}$ & $22\pm5$ & $23^{+5}_{-3}$  & $20\pm5$  \\
$\BR_{\phiy}^{\eta\phi}\Gamma^{\EE}_{\phiy} (\ev)$
                & $0.48^{+0.04}_{-0.02}$ & $0.37\pm 0.03$ & $38\pm 1$   & $37^{+2}_{-1}$   \\
$M_{\phiy} (\mevcs)$    & \multicolumn{4}{c}{$2162(fixed)$} \\
$\Gamma_{\phiy} (\mev)$ & \multicolumn{4}{c}{$100(fixed)$}   \\
$\theta_{\phit}(^\circ)$ & $80\pm12$ &  $-109\pm 11$ & $-88^{+12}_{-7}$ & $-54\pm6$   \\
$\theta_{\phiy}(^\circ)$ & $-61^{+14}_{-10}$  & $-48^{+10}_{-8}$ & $-173^{+16}_{-12}$ & $-165\pm 9$  \\
\hline \hline
\end{tabular}
\label{fit_results1}
\end{center}
\end{table}

As discussed in Ref.~\cite{zhukai}, there are $2^{n-1}$ solutions in a fit with $n$ components in the 
amplitude. They have the same goodness of fit, and the same mass and width of a resonance.  Unfortunately, we can not 
find a proliferation of the solutions, like in some previous measurements~\cite{y2175_belle, paper_1, paper_3, 
paper_4}.

\subsection{Fit with the $\phit$, $X(1750)$, $\phiy$ and a non-resonant components}

\begin{figure}[tbp]
\centering
\includegraphics[width=0.4\textwidth]{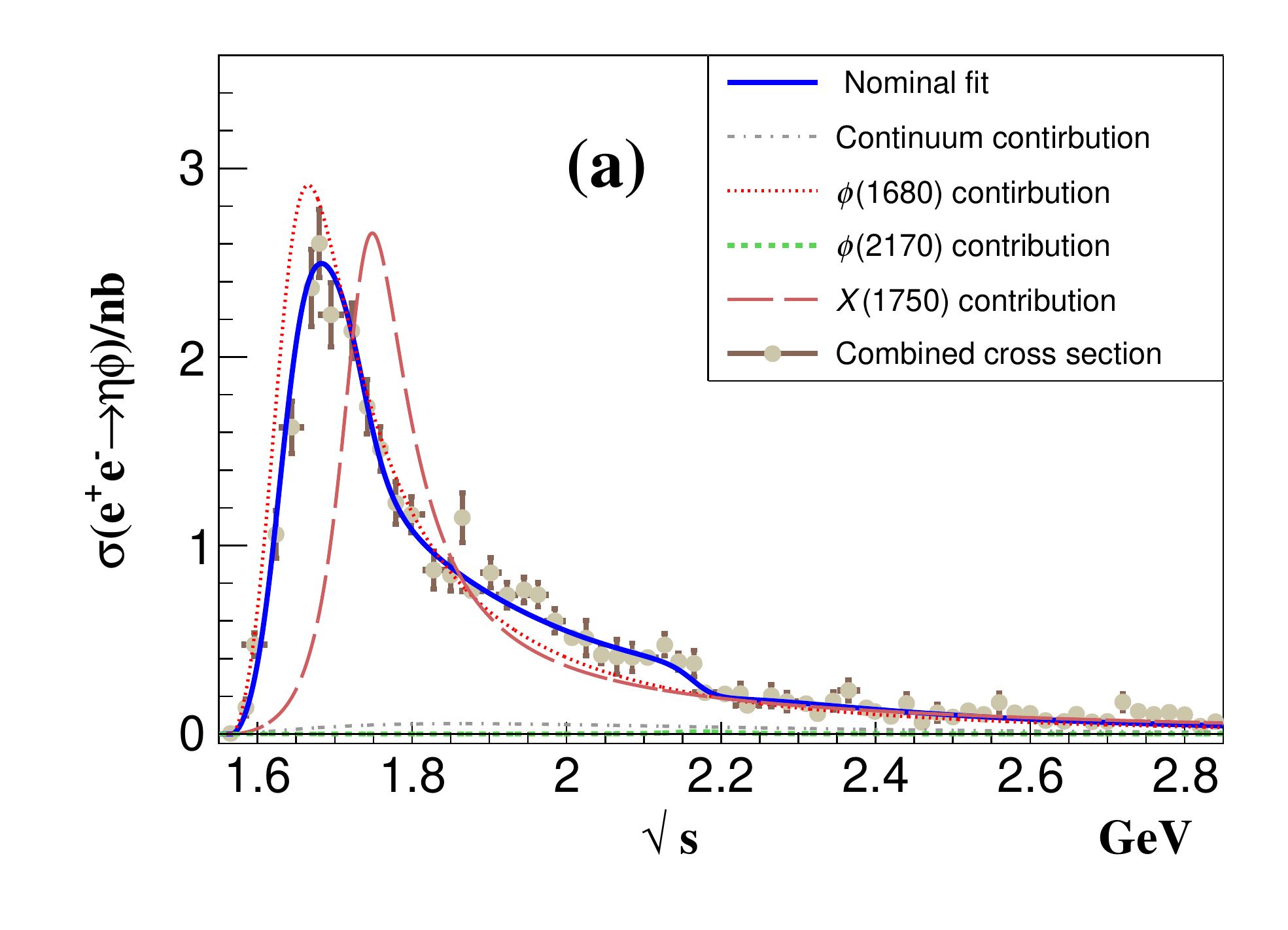}
\includegraphics[width=0.4\textwidth]{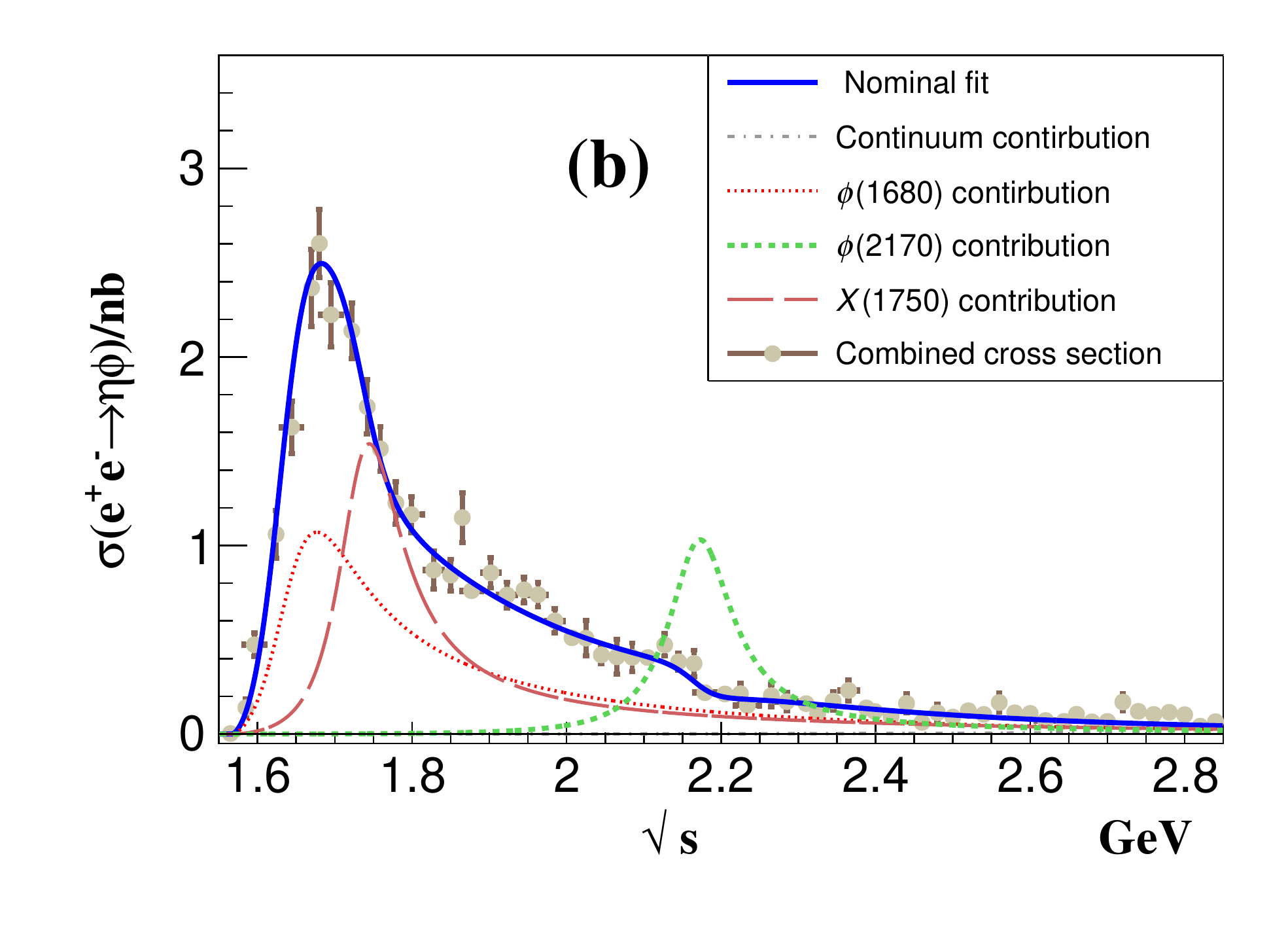}
\caption{Two solutions of fitting to the $\sigma(\EE\to\eta\phi)$ measured by the BaBar, Belle, BESIII and CMD-3 
experiments with $\phit$, $X(1750)$ and $\phiy$. The blue solid lines show the best fit results, and the dashed red, 
orange, green and gray lines show the $\phit$, $X(1750)$, $\phiy$ and non-resonant components, respectively. The 
interference among the $\phit$, $X(1750)$, $\phiy$ and non-resonant components are not shown.}
\label{fit2}
\end{figure}

To investigate the production of $X(1750)$ in the $\EE\to\eta\phi$ process, we perform the combined fit with the 
Eq.~\ref{eq_2}. We fix the mass and width of $X(1750)$ to the world average values~\cite{PDG}. We obtain eight 
solutions of equivalent quality of $\chi^2/ndf = 290/245$, having the same masses and widths of $\phit$ and $\phiy$. 
The fit results are listed in Table~\ref{fit_results2} and the first two solutions are shown in Fig.~\ref{fit2}. 
Comparing $\Delta\chi^2 = 6$ and $\Delta ndf = 2$ in the fits with and without $X(1750)$, the statistical 
significance of the $X(1750)$ is $2.0\sigma$. Since the $X(1750)$ is not significant here, we determine the upper 
limits (UL) of its production ($\BR_{X(1750)}^{\eta\phi}\Gamma^{\EE}_{X(1750)}$) in the eight solutions at 90\% C.L. 
by integrating the likelihood versus the $X(1750)$ yield, as listed in Table~\ref{fit_results2}.  

\begin{sidewaystable}[htbp]
\begin{center}
\caption{Results of fitting to the $\sigma(\EE\to\eta\phi)$ measured by BaBar, Belle, BESIII and CMD-3 experiments 
with $\phit$, $X(1750)$, $\phiy$ and non-resonant components. The mass and width of $X(1750)$ are fixed to the world 
average values~\cite{PDG}.}
\begin{tabular}{c|c c c c c c c c}\hline\hline
Parameters   &  Solution I   & Solution II   & Solution III    & Solution IV  & Solution V & Solution VI & Solution VII & Solution VIII  \\\hline
$\chi^2/ndf$ & \multicolumn{8}{c}{290/245}   \\
$a_{0}$      & $2.2^{+0.4}_{-0.2}$ & $-1.2\pm 0.2$ & $-0.15\pm 0.04$ & $-0.12\pm 0.02$ & $-0.62^{+0.04}_{-0.03}$ & $0.14\pm 0.04$ & $1.4^{+0.3}_{-0.2}$ & $-1.2\pm 0.4$ \\
$a_{1}$      & $2.9^{+0.3}_{-0.2}$ & $2.7\pm 0.3$ & $0.61^{+0.04}_{-0.02}$ & $-1.2^{+0.2}_{-0.1}$ & $1.6\pm 0.3$ & $0.79\pm 0.04$ & $1.6\pm 0.2$ & $2.5\pm 0.2$ \\
$\BR^{\eta\phi}_{\phit}\Gamma^{\EE}_{\phit} (\ev)$ 
             & $247^{+9}_{-7}$ & $93 \pm 6$ & $107 \pm 9$ & $159^{+11}_{-7}$ & $244^{+9}_{-5}$ & $170^{+11}_{-7}$ & $114^{+9}_{-5}$ & $280\pm 12$\\
$M_{\phit} (\mevcs)$   & \multicolumn{8}{c}{$1680 \pm 4$}   \\
$\Gamma_{\phit}(\mev)$ & \multicolumn{8}{c}{$147 \pm 8$}   \\
$\BR^{\eta\phi}_{\phit}(\%)$  
             & $19\pm 3$ & $18\pm 3$ & $19^{+4}_{-3}$ & $21\pm3$ & $22\pm 3$ & $19\pm 3$ & $21^{+4}_{-2}$ & $22\pm4$\\
$\BR_{X(1750)}^{\eta\phi}\Gamma^{\EE}_{X(1750)} (\ev)$ 
             & $210^{+18}_{-14}$ & $102^{+15}_{-12}$ & $167^{+22}_{-16}$ & $172\pm 19 $ & $227^{+20}_{-16}$ & $250^{+23}_{-17}$ & $289\pm22$ & $102^{+18}_{-14}$ \\
UL of $\BR_{X(1750)}^{\eta\phi}\Gamma^{\EE}_{X(1750)} (\ev)$ & 249 & 136 & 197 & 214 &  269 &  287 & 322 & 142 \\
$M_{X(1750)} (\mevcs)$    & \multicolumn{8}{c}{ 1754 (fixed)} \\
$\Gamma_{X(1750)} (\mev)$ & \multicolumn{8}{c}{ 120 (fixed)}   \\                
$\BR_{\phiy}^{\eta\phi}\Gamma^{\EE}_{\phiy} (\ev)$ & $0.34 \pm 0.02$ & $38^{+2}_{-1}$ & $0.57 \pm 0.04$   & $39^{+2}_{-1}$ & $0.42 \pm 0.02$ & $37^{+2}_{-1}$ 
			& $0.44^{+0.04}_{-0.02}$ & $41\pm 2$\\
$M_{\phiy} (\mevcs)$    & \multicolumn{8}{c}{$2169^{+8}_{-6}$} \\
$\Gamma_{\phiy} (\mev)$ & \multicolumn{8}{c}{$95^{+22}_{-14}$}   \\
$\theta_{\phit} (^\circ)$ & $98^{+11}_{-9}$ & $109 \pm 17$ & $88 \pm 16$ & $-97^{+11}_{-9}$ &  $-134 \pm 17$ &$119 \pm 15$ & $-125 \pm 19$ 
			& $-109^{+20}_{-14}$ \\
$\theta_{X(1750)}(^\circ)$ & $-55 \pm 7$ & $-68^{+18}_{-12}$ & $-74 \pm 14$ & $105^{+20}_{-14}$ & $63 \pm 12$ & $-59 \pm 14$ & $108 \pm 15$ & $131^{+18}_{-13}$ \\
$\theta_{\phiy)}(^\circ)$  & $-118 \pm 15$ & $-94^{+16}_{-13}$ & $-108 \pm 14$ & $132 \pm 21$ & $-83 \pm 17$ & $-69^{+15}_{-11}$ & $-127 \pm 24$ & $-111 \pm 17$\\
\hline\hline
\end{tabular}
\label{fit_results2}
\end{center}
%\end{table}
\end{sidewaystable}

\section{Systematic uncertainties}
\label{sys_err}

We characterize the following systematic uncertainties for the nominal fit results. We estimate the uncertainty of 
parametrization in Eq.~\ref{eq_3} with two different parametrization methods, which have the forms
\beq\label{eq_11}
A(M)\propto \frac{\sqrt{\Gamma_{\phit}(\sqrt{s})\Gamma_{\EE}}}{s-M^{2}+iM\Gamma_{\phit}}
\eeq
and
\beq\label{eq_10}
A(M)\propto \frac{M}{\sqrt{s}}\cdot\frac{\sqrt{\Gamma_{\phit}(\sqrt{s})\Gamma_{\EE}}}{s-M^{2}+iM\Gamma_{\phit}}.
\eeq
By changing the fit range to $[1.6,~2.9]~\gevcs$, we find the systematic uncertainty due to the fit range is 
negligible. We use $A_{\eta\phi}^{n.r.}(s) = a_{0}/s$ to estimate the model dependence of the non-resonant 
contribution. We obtain the uncertainty in $\BR^{\kkt}_{\phit}/\BR^{\eta\phi}_{\phit}$ by varying $1\sigma$ according 
to the previous measurement~\cite{etaphi_babar}. To estimate the uncertainty due to the possible contribution from 
$\phit\to\phi\pi\pi$, we take $\BR(\phit\to \phi\pi\pi) = \BR(\phit\to \phi\eta)/2$~\cite{y2175_belle} and modify the 
Eq.~(\ref{eq_4}) to 
\beqar\label{eq_12}
\Gamma_{\phit}(\sqrt{s}) & = &
\Gamma_{\phit}\cdot [\frac{\mathcal{P}_{KK^{*}(892)}(\sqrt{s})}{\mathcal{P}_{\kkt}(M_{\phit})}\BR^{\kkt}_{\phit} +
\frac{\mathcal{P}_{\eta\phi}(\sqrt{s})}{\mathcal{P}_{\eta\phi}(M_{\phit})}\BR^{\eta\phi}_{ \phit} \nonumber \\
& & +\frac{\mathcal{P}_{\phi\pi\pi}(\sqrt{s})}{\mathcal{P}_{\phi\pi\pi}(M_{\phit})}\BR^{\phi\pi\pi}_{ \phit}
 + (1-\BR^{\eta\phi}_{\phit}-\BR^{\phi\pi\pi}_{\phit}-\BR^{\kkt}_{\phit})]
\eeqar
in the combined fits. Here, $\mathcal{P}_{\phi\pi\pi}$ is the phase space of the $\phit\to \phi \pi\pi$ decay. 

Assuming all these sources are independent and adding them in quadrature, the total systematic uncertainties are 
listed in Table.~\ref{sys_err}.

\begin{table}[htbp]
\begin{center}
\caption{Systematic uncertainties of the resonances parameters for $\phit$ and $\phiy$. $M$, $\Gamma$, 
$\Gamma^{\EE}\BR$ and $\BR$ are the mass with unit $\mevcs$, total width with unit $\mev$, the production of the 
branching fraction and the partial width to $\EE$ with unit $\ev$, and the branching fraction (\%). }
\resizebox{\textwidth}{!}{
\begin{tabular}{ c | c | c | c | c | c | c }\hline\hline
\diagbox{Parameter}{Source} & & $A_{\eta\phi}^{n.r.}(s)$ & $\BR_{\phit}^{\kkt}/\BR_{\phit}^{\eta\phi}$ & Parametrization & $\phit\to\phi\pi\pi$ & Sum \\\hline
$M_{\phit}$ 	& & 2 & 3 & 3  & 5 & 7 \\
$\Gamma_{\phit}$ 	&  & 3 & 5 & 5 & 4 & 9 \\\hline
\multirow{4}{*}{$\Gamma_{\phit}^{\EE}\BR_{\phit}^{\eta\phi}$} & Sol. I & 8 & 10  & 7 & 6 & 16 \\
& Sol. II 	& 4 & 9 & 6 & 2 & 12 \\
& Sol. III 	& 7 & 5 & 7 & 6 & 13 \\
& Sol. IV 	& 6 & 5 & 4 & 7 & 11 \\\hline
\multirow{4}{*}{$\BR_{\phit}^{\eta\phi}$} & Sol. I & 0.7 & 0.9 & 1.3 &  1 & 2 \\
& Sol. II 	& 0.5 & 0.9 & 1.2 & 2 & 4 \\
& Sol. III 	& 0.9 & 1.5 & 1.8 & 2 & 3 \\
& Sol. IV 	& 0.7 & 0.4 & 1.8 & 1 & 3 \\\hline
$M_{\phiy}$ & 	& 3 & 5 & 2   & 2 & 6\\
$\Gamma_{\phiy}$ 	& & 6 & 2 & 6 & 3 & 9 \\\hline
\multirow{4}{*}{$\Gamma_{\phiy}^{\EE}\BR_{\phiy}^{\eta\phi}$} & Sol. I & 0.02 & 0.03 & 0.05 & 0.03 & 0.07 \\
& Sol. II 	& 0.02 & 0.04 & 0.04 & 0.03 & 0.07 \\
& Sol. III 	& 2 & 2 & 3 & 2 & 5 \\
& Sol. IV 	& 3 & 3 & 4 & 1 & 6 \\
\hline\hline
\end{tabular}}
\label{sys_err}
\end{center}
\end{table}

\section{Summary}

Combining the measurements by the BaBar, Belle, BESIII and CMD-3 experiments, we calculate the 
$\sigma(\EE\to\eta\phi)$ from threshold to $3.95~\gev$ with an improved precision. There are clear $\phit$ and 
$\jpsi$ signals and lineshape changes around the mass of $\phiy$ in the $\eta\phi$ final state. We perform combined 
fits to $\sigma(\EE\to\eta\phi)$ measured by the four experiments and get the nominal fit results with $\phit$, 
$\phiy$ and non-resonant components. The statistical significance of $\phiy$ is $7.2\sigma$. The mass and width of 
$\phiy$ are $M_{\phiy} = (2169 \pm 5 \pm 6)~\mevcs$, $\Gamma_{\phiy} = (96^{+17}_{-14}\pm 9)~\mev$, which are 
consistent with the world average values~\cite{PDG}. The mass and width of $\phit$ are $M_{\phit} = (1678^{+5}_{-3} 
\pm 7)~\mevcs$ and $\Gamma_{\phit} = (156\pm 5 \pm 9)~\mev$, with a good precision comparing to the world average 
values~\cite{PDG}. The branching fraction of $\phit\to\eta\phi$ decay is about 20\% with uncertainties less than 6\%. 
We also determine the $\BR_{\phit}^{\eta\phi}\Gamma^{\EE}_{\phit}$ and $\BR_{\phiy}^{\eta\phi}\Gamma^{\EE}_{\phiy}$ 
from the fits. Assuming the existence in the $\EE\to\eta\phi$ process, the statistical significance of $X(1750)$ is 
only $2.0\sigma$. We determine the upper limit of $X(1750)$ in $\EE\to\eta\phi$ at 90\% C.L. 

\acknowledgments

We thank Prof. Changzheng Yuan for very helpful discussions. This work is supported by National Key R\&D 
Program of China under Contract No.~2022YFA1601903 and the National Natural Science Foundation of China under 
Contract No. 12175041.


\begin{thebibliography}{**}

\bibitem{babay4260} B.~Aubert {\it et al.} (BaBar Collaboration),
\Journal\PRL{95}{142001}{2005}.

\bibitem{y2175_babar} B.~Aubert {\it et al.} (BaBar Collaboration),
\Journal\PRD{74}{091103}{2006}.

\bibitem{y2175_belle} C.~P~Shen {\it et al.} (Belle Collaboration),
\Journal\PRD{80}{031101}{2009}.

%\bibitem{PDG} P.A. Zyla {\it et al.} (Particle Data Group),
%Prog. Theor. Exp. Phys. {\bf 2020}, 083C01(2020) and 2021 update.
\bibitem{PDG} R.L. Workman {\it et al.} (Particle Data Group),
\Journal\PTEP{2022}{083C01}{2022}.

\bibitem{etaphi_babar} B.~Aubert {\it et al.} (BaBar Collaboration),
\Journal\PRD{76}{092005}{2007}.

\bibitem{etaphi_babar_2} B.~Aubert {\it et al.} (BaBar Collaboration),
\Journal\PRD{77}{092002}{2008}.

\bibitem{etaphi_belle} W. J.~Zhu {\it et al} (Belle Collaboration),
\Journal\PRD{107}{012006}{2023}.

\bibitem{etaphi_cmd3} V.~L.~Ivanov {\it et al.},
\Journal\PLB{798}{134946}{2019}.

\bibitem{etaphi_bes3} M.~Ablikim {\it et al} (BESIII Collaboration),
\Journal\PRD{104}{032007}{2021}.

\bibitem{etapphi_bes3} M.~Ablikim {\it et al.} (BESIII Collaboration),
\Journal\PRD{102}{012008}{2020}.

\bibitem{lattice-qcd} Y. H.~Ma, Y.~Chen, M.~Gong and Z. F.~Liu,
\Journal\CPC{45}{013112}{2021}.
%arXiv:2007.14893v1.


\bibitem{strange} T.~Barnes, N.~Black and R.~R.~Page,
\Journal\PRD{68}{054014}{2003}.

\bibitem{FOCUS} J.~M.~Link {\it et al.},
\Journal\PLB{545}{50}{2002}.

\bibitem{kketa} M.~Ablikim {\it et al} (BESIII Collaboration),
\Journal\PRD{101}{032008}{2020}.

\bibitem{Schmelling} M. Schmelling, 
\Journal\PS{51}{1995}{676}.

%\bibitem{Belle} A.~Abashian {\it et al.} (Belle Collaboration),
%A.~Abashian {\it et al.} (Belle Collaboration),
%\Journal\NIMA{479}{117}{2002};
%also see detector section in J.Brodzicka {\it et al.},
%\Journal\PTEP{2012}{04D001}{2012}.
%% \Prog. Theor. Exp. Phys. {\bf 2012}, 04D001 (2012).

%\bibitem{KEKB} S.~Kurokawa and E.~Kikutani, Nucl.
%\Journal\NIMA{499}{1}{2003}, and other papers included in this Volume;
% T.Abe {\it et al.},
% \Journal\PTEP{2013}{03A001}{2013} and references therein.

\bibitem{minuit} F. James and M. Roos,
\Journal\CoPC{10}{343}{1975}.
%Comput. Phys. Commun. 10, 343 (1975).

\bibitem{zhukai} K.~Zhu, X.~H.~Mo, C.~Z.~Yuan and P.~Wang,
\Journal\IJMP{26}{2011}{4511}.

\bibitem{paper_1}C.~Z.~Yuan {\it et al.}, Belle Collaboration,
\Journal\PRL{99}{182004}{2007}.

\bibitem{paper_3} X.~L.~Wang {\it et al.}, Belle Collaboration,
\Journal\PRD{91}{112007}{2015}.

\bibitem{paper_4}X.~L.~Wang {\it et al.}, Belle Collaboration,
\Journal\PRD{87}{051101}{2013}.

%%%%%%%%%%%%%%%%%%%%%%%%%%%%
\end{thebibliography}
\end{document}